\documentclass[pre,aps,twocolumn,amsmath,amssymb,showpacs]{revtex4}
\usepackage{amssymb}
\usepackage{graphicx}
\usepackage{hyperref}
\usepackage{float}
\usepackage[utf8x]{inputenc}
\usepackage{gensymb}
\usepackage{color}
\usepackage[utf8x]{inputenc}
\begin{document}

\title{Adsorption of 2d polymers with two- and three-body self-interactions}

\author{Nathann T. Rodrigues}\email{nathan.rodrigues@ufv.br}
\affiliation{Departamento de F\'isica, Universidade Federal de Vi\c cosa, 36570-900, Vi\c cosa, MG, Brazil}
\author{Tiago J. Oliveira}\email{tiago@ufv.br}
\affiliation{Departamento de F\'isica, Universidade Federal de Vi\c cosa, 36570-900, Vi\c cosa, MG, Brazil}
\author{Thomas Prellberg}\email{t.prellberg@qmul.ac.uk}
\affiliation{School of Mathematical Sciences, Queen Mary University of London, Mile End Road, London, E1 4NS, United Kingdom}
\author{Aleksander L. Owczarek}\email{owczarek@unimelb.edu.au}
\affiliation{School of Mathematics and Statistics, University of Melbourne, Victoria 3010, Australia}

\date{\today}

\begin{abstract}
Using extensive Monte Carlo simulations, we investigate the surface adsorption of self-avoiding trails on the triangular lattice with two- and three-body on-site monomer-monomer interactions.
In the parameter space of two-body, three-body, and surface interaction strengths, the phase diagram displays four phases: swollen (coil), globule, crystal, and adsorbed.
For small values of the surface interaction, we confirm the presence of swollen, globule, and crystal bulk phases. For sufficiently large values of the surface interaction, the system is in an adsorbed state, and the adsorption transition can be continuous or discontinuous, depending on the bulk phase. 
As such, the phase diagram contains a rich phase structure with transition surfaces that meet in multicritical lines joining in a single special multicritical point.
The adsorbed phase displays two distinct regions with different characteristics, dominated by either single or double layer adsorbed ground states. Interestingly, we find that there is no finite-temperature phase transition between these two regions though rather a smooth crossover.
\end{abstract}


\maketitle

\section{Introduction}
\label{intro}

Understanding the behavior of self-interacting single polymer chains in a solvent is of fundamental interest from both theoretical and experimental point of view \cite{DesCloizeaux1990,DeGennes1979}. In general, one can find a rich phase structure. In a good solvent, a polymer exists in a swollen globule state, whereas in a bad solvent the polymer collapses into a partially dense amorphous globule or presents in a fully dense crystal-like structure. 

The study of a variety of lattice models of polymers has played a fundamental role in elucidating these phases and phase transitions between them. For example, the coil-globule transition between a swollen coil and a collapsed globule at the so-called $\theta$-point has been analyzed in both two and three dimensions \cite{DeGennes1975,Stephen1975,Duplantier1982}, with theoretically predicted critical exponents \cite{Duplantier1986,Duplantier1987,Duplantier1987a} associated with the transition confirmed by simulations and experiment.

\begin{figure}[!b]
\includegraphics[width=7.5cm]{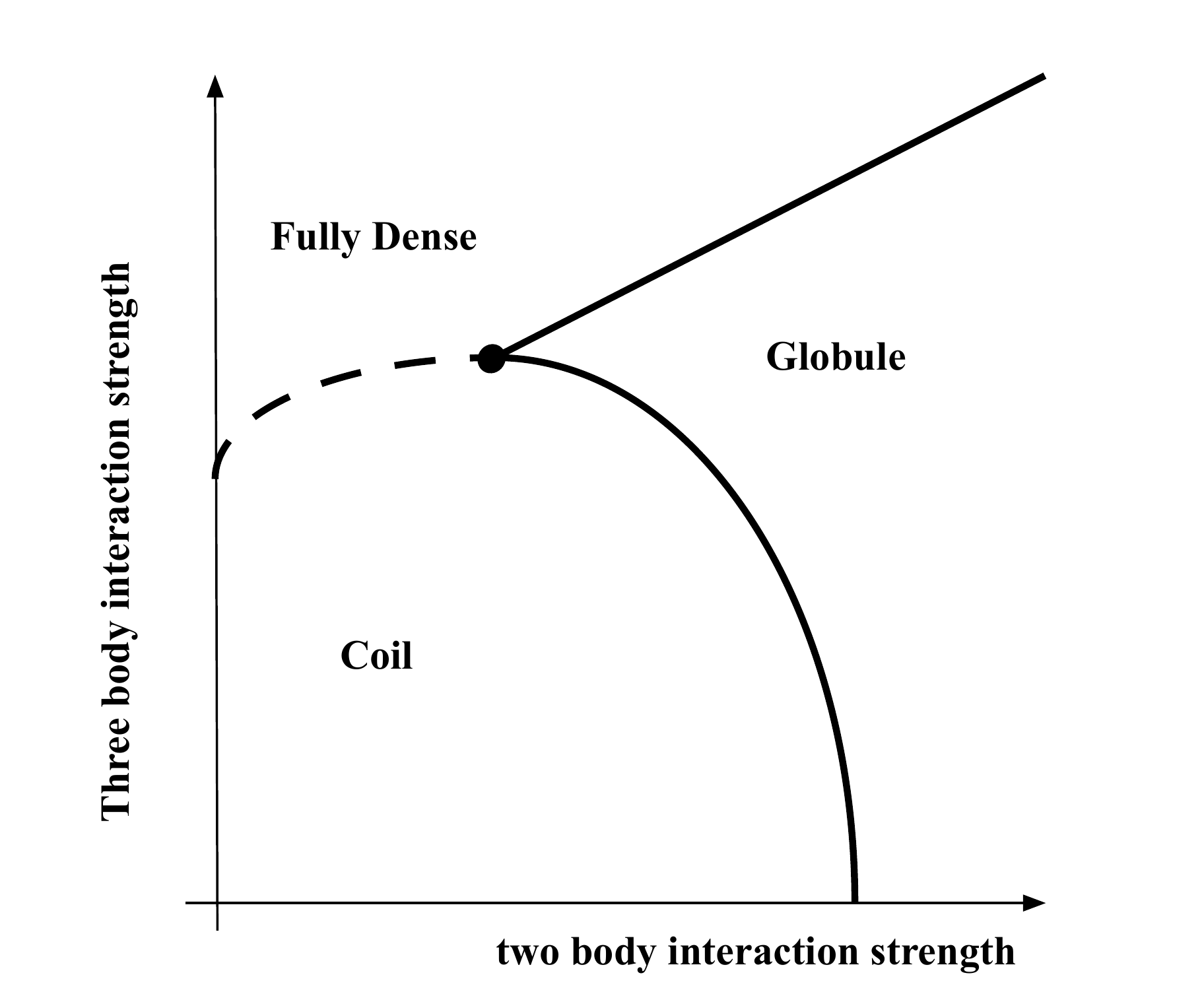}
\caption{Schematic phase diagram for a single polymer chain in a solvent with two-body and three-body interactions, showing the presence of coil, globule, and fully dense phases. The coil-globule transition is critical, whereas the coil-dense transition is first-order. In two dimensions, the globule-dense transition is expected to be continuous.}
\label{Fig0}
\end{figure}

More generally, interacting self-avoiding walks and trails with competing two-body and three-body interactions give rise to phase diagrams which in addition to the swollen phase show both collapsed globule and dense crystal-like phases, with all three phases meeting at a multi-critical point \cite{Doukas2010,Bedini2012,Bedini2016}. A schematic phase diagram is shown in Fig.~\ref{Fig0}.

Adsorption of a polymer chain onto a surface is another fundamental problem in polymer physics \cite{Eisenriegler1982,DeBell1993,Vrbova1996,Vrbova1998,Vrbova1999,Grassberger2005,Owczarek2007,Luo2008,Klushin2013}. If a polymer is tethered with one end to a sticky surface, then increasing the strength of the surface attraction induces an adsorption transition from a desorbed bulk state to an adsorbed state in which the polymer is bound to the surface. If the polymer is in a swollen bulk phase, then one speaks of a ``normal'' surface transition, whereas if the polymer in bulk is in the critical $\theta$-state, one speaks of a ``special'' surface transition, each of which is associated with its own set of critical surface exponents~\cite{Nathann2019}. If the polymer adsorbs from a collapsed coil, one finds a surface-attached globule (which is absent in two-dimensional models, in which the adsorbing surface is modeled by a line).

Recently, there has been renewed interest in the polymer adsorption transition, due to the fact that numerical studies pointed at possible non-universal behavior of the normal adsorption transition \cite{Plascak2017,Martins2018,Bradly2018,Bradly2018a}. Of interest here is the fact that these works show that the determination of these exponents is rather subtle due to the presence of strong finite-size effects, despite the fact that the phases themselves could be identified in a robust manner.

In this paper, we consider the adsorption transition of a polymer chain with both two-body and three-body interactions. The lattice model considered here is one of self-avoiding trails with on-site interactions on the triangular lattice, which has been studied previously \cite{Doukas2010}. We now study this model in a half-plane with the trails restricted to start at the origin, and adsorption mediated by on-site attractive interactions at vertices in the boundary line. In addition to the three bulk phases, we find an adsorbed phase. All phases meet each other pairwise at transition surfaces, which meet each other in multicritical lines joining in a single special multicritical point. The analysis is made somewhat more difficult by the fact that the adsorbed phase shows two distinct regions with different characteristics, dominated by either single or double layer adsorbed ground states.

This paper is organized as follows. In Section \ref{defmod} we define the model and the quantities used to investigate its thermodynamics properties. Some details on the Monte Carlo methods and simulations are given in Section \ref{Simul}. Sections \ref{boundary} and \ref{3d} present results respectively for the boundary planes and for slices of the three-dimensional parameter space. A summary of the full phase diagram is presented in Section \ref{SecConclusions}.

\section{Model and quantities of interest}
\label{defmod}

A self-avoiding trail (SAT) is defined as a path on a regular lattice, composed of a finite collection of adjacent vertices on the sites of the lattice that are linked by steps along the edges of the lattice, with the restriction that all edges have to be unique. We identify the vertices of this path as monomers and the edges as a sequence of bonds connecting such monomers, in a way that no closed loop is formed. Thereby, the polymer excluded volume interaction is introduced in this model by the restriction of one bond per edge, while more than one monomer can be placed at the same site. In a lattice of coordination number $q$, the maximal number of monomers at one site is $\lfloor q/2\rfloor$. 

\begin{figure*}[!t]
\includegraphics[width=11.50cm]{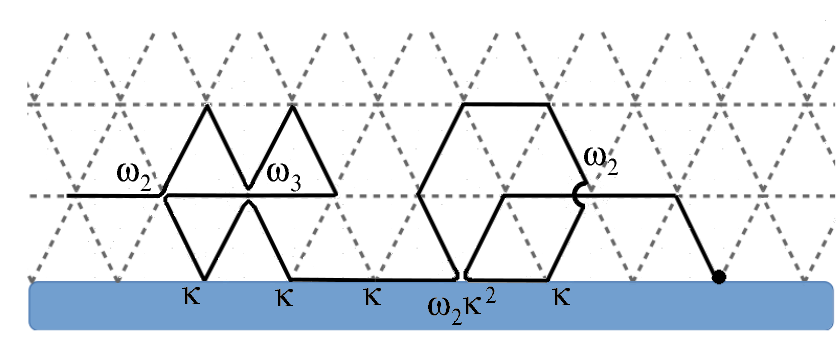}
\caption{Illustration of a self-avoiding trail with $n=22$ steps on the triangular lattice. The solid black circle denotes the origin where the trail is tethered. The Boltzmann weights associated with different site configurations are also indicated.}
\label{Fig1}
\end{figure*}

Here, we are interested in investigating transitions among bulk and adsorbed phases of polymers modeled as self-attracting SATs defined on the half-plane consisting of a triangular lattice limited by a horizontal surface (see Fig.~\ref{Fig1}). As the triangular lattice has coordination number $q=6$, bulk sites can be visited by up to three monomers. The coordination number of sites on the surface is $4$, so surface sites can have at most two monomers. Self-attraction is included in the SATs by associating energies $-\epsilon_2 \leqslant 0$ to doubly visited (regardless they are in surface or bulk) and $-\epsilon_3 \leqslant 0$ to triply visited sites. Moreover, a polymer-surface interaction is introduced by assigning an energy $-\epsilon_s \leqslant 0$ to each monomer lying on the surface. A configuration for this system is illustrated in Fig.~\ref{Fig1}. The trail is tethered to the origin, which is located on the surface. 

In order to write down the partition function of the system, we associate Boltzmann weights $\omega_2=e^{\beta\epsilon_2}$ and $\omega_3=e^{\beta\epsilon_3}$ to each doubly and triply visited site, respectively, and $\kappa=e^{\beta\epsilon_s}$ to each monomer on the surface, with $\beta = 1/(k_B T)$, where $k_B$ is the Boltzmann constant and $T$ the temperature. The partition function is then given by
\begin{equation}
 Z_n(\omega_2,\omega_3,\kappa) = \sum_{m_2,m_3,m_s} C^{(n)}_{\{m_2,m_3,m_s\}} \omega_2^{m_2} \omega_3^{m_3} \kappa^{m_s},
\end{equation}
where $C^{(n)}_{\{m_2,m_3,m_s\}}$ is the number of $n$-step lattice trails with $m_2$ ($m_3$) doubly (triply) visited sites and $m_s$ monomers on the surface. The expected value of any thermodynamic quantity $Q$ is then defined by the average
\begin{equation}
 \left<Q\right>(\omega_2,\omega_3,\kappa) = \frac{1}{Z_n}\sum_{\psi_n}\omega_2^{m_2(\psi_n)}\omega_3^{m_3(\psi_n)}\kappa^{m_s(\psi_n)}Q(\psi_n),
\end{equation}
where the sum runs over all $n$-step trails $\psi_n$. We are particularly interested in the energy contribution per step, $e_j^{(n)}$, for doubly, triply and surface contacts:
\begin{equation}
  e_j^{(n)} = \frac{\left<m_j\right>}{n} = \frac{1}{n Z_n}\sum_{m_2,m_3,m_s} m_j C^{(n)}_{\{m_2,m_3,m_s\}} \omega_2^{m_2} \omega_3^{m_3} \kappa^{m_s},
\end{equation}
where $j=2,3,s$. 

A key quantity to identify possible phase transitions is the fluctuation, $c_j^{(n)}$, of these energy contributions:
\begin{equation}
 c_j^{(n)}(\omega_2,\omega_3,\kappa)=\frac{\left<m_j^2\right>-\left<m_j\right>^2}{n}.
\end{equation}
Close to a continuous bulk transition the fluctuations $c_2^{(n)}$ and $c_3^{(n)}$ of doubly and triply visited sites, respectively, are expected to follow the scaling law 
\begin{equation}
 c_{i}^{(n)}\sim n^{{2\phi_b}-1}h(\tau n^{\phi_b}),
 \label{BulkF}
\end{equation}
where $\phi_b$ is the crossover exponent related to bulk transitions, $\tau\equiv T - T_c$ is the temperature relative to the bulk transition temperature $T_c$, and $h(\cdot)$ is a scaling function. For an adsorption transition, the internal surface energy $e_s^{(n)}$ is the order parameter. Close to the adsorption point the surface energy is expected to behave as
\begin{equation}
 e_s^{(n)}\sim n^{\phi_s - 1}f(\tau n^{1/\delta}),
\end{equation}
where $\tau\equiv T - T_a$ is the temperature relative to the adsorption temperature $T_a$, $\phi_s$ is a critical exponent, $1/\delta$ is the crossover exponent associated with the adsorption transition and $f(\cdot)$ is a scaling function.

In order to analyze the phase transitions, we need to introduce several quantities. The exponent $1/\delta$ can be obtained from the quantity
\begin{equation}
 \Gamma_n=\frac{\left<m_s^2\right>-\left<m_s\right>^2}{\left<m_s\right>},
 \label{Gn}
\end{equation}
whose maximum in curves of $\Gamma_n$ as function of $T$ behaves as
\begin{equation}
 \Gamma_{n,max}\sim n^{1/\delta}.
 \label{Gn_es}
\end{equation}

Although we will not be concerned with a careful study of critical exponents here, in some cases estimates for such exponents are very important to locate the transition points. In particular, the crossover exponents $\phi_b$ and $1/\delta$ are crucial if we want to determine $T_{x,\infty}$ from finite-size estimates $T_{x,n}$, for $x=c$ or $a$, since
\begin{equation}
 T_{x,n} \simeq T_{x,\infty} + n^{-\psi},
 \label{Tn}
\end{equation}
where $\psi$ is equal to $\phi_b$ (when $x=c$) or $1/\delta$ (when $x=a$).

Some metric quantities, such as the mean squared end-to-end distance, $R^2_n$, are also important in the study of both bulk and surface transitions. For our simulations we have created a triangular lattice by augmenting a square lattice with diagonals, and a simple linear transform implies that the parallel and perpendicular components of $R^2_n$ with respect to the surface are given by
\begin{align}
    R^2_{\perp,n} (\omega_2,\omega_3,\kappa)&= \frac34\langle y_n^2 \rangle, \\
    R^2_{\parallel,n} (\omega_2,\omega_3,\kappa)&= \langle x_n^2 \rangle + \frac14\langle y_n^2 \rangle + \langle x_n y_n \rangle.
    \label{eq:EndToEndRadius}
\end{align}
Close to the adsorption such quantities are expected to follow
\begin{equation}
  R^2_{\perp/\parallel,n} \sim n^{2 \nu_{\perp/\parallel}} g(\tau n^{1/\delta}),
\end{equation}
where $g(\cdot)$ is a scaling function and $\nu_{\perp/\parallel}$ are the Flory exponents. Clearly the scaling of $R^2_n=R^2_{\perp,n}+R^2_{\parallel,n}$ is dominated by the scaling of the larger quantity on the right-hand-side, and thus has an associated Flory exponent $\nu=\max(\nu_{\perp},\nu_{\parallel})$.
In non-adsorbed (bulk) phases, such exponents are equal (i.e., $\nu_\parallel=\nu_\perp$) in the thermodynamic limit, but in finite systems one finds effective exponent estimates $\nu_{\perp/\parallel,n}$ for which $\nu_{\parallel,n} < \nu_{\perp,n}$. In adsorbed phases, on the other hand, the trails behave as quasi one-dimensional walks, so that $\nu_{\parallel,n}\rightarrow 1$ and $\nu_{\perp,n}\rightarrow 0$. Therefore, in finite trails these exponents cross at some temperature, $T_{a,n}$, located in between the non-adsorbed and adsorbed phases. These crossing temperatures can thus be used as finite-size estimate of the adsorption transition temperature.

We can also use the exponent $\nu$ to locate bulk transitions such as the coil-globule transition. In two dimensions,
in the coil phase $\nu$ assumes the value of $3/4$~\cite{Flory1971}, while in the globule phase a value $\nu=1/2$~\cite{Flory1971} is expected. For finite systems, however, we have that $\nu_{coil,n} \rightarrow 3/4$ from below and $\nu_{globule,n} \rightarrow 1/2$ from above as the length $n$ increases, so that curves of $\nu_n$ against $T$ for different values of $n$ cross at some intermediate temperature, which can be taken as an estimate of $T_{c,n}$. At criticality, we expect to find $\nu=4/7$~\cite{Duplantier1987}.
In a similar fashion, we can locate the collapse transition by using the scaling of the partition function
\begin{equation}
 Z_n \sim \mu^n n^{\gamma - 1},
\end{equation}
where $\mu$ is the connective constant and $\gamma$ the entropic exponent. Similarly to $\nu$, curves of $\gamma \times T$ for different lengths $n$ should cross each other inbetween the coil and globule phases, giving also estimates of $T_{c,n}$.

\section{Numerical Simulations}
\label{Simul}

In order to build up the full phase diagram of the model for a broad range of parameters, we sample SATs with the flatPERM algorithm~\cite{Prellberg2004}. Similarly to the pruned-enriched Rosenbluth method (PERM)~\cite{Grassberger1997}, flatPERM is an improvement of the classical Rosenbluth algorithm. In both flatPERM and PERM, a SAT is stochastically grown with pruning and enrichment strategies intended to prevent its attrition and the appearance of rare configurations with small or extremely large statistical weights. In the flatPERM, the pruning and enrichment processes are implemented independently of the thermodynamic parameters of the system, so that the method gives us an estimate of the density of states $C^{(n)}_{\{m_2,m_3,m_s\}}$. PERM, on the other hand, gives us an estimate of the partition function $Z_n$ for a given set of parameters.

The dimensionality of the density of states --- i.e., the number of energy parameters in the model, which is three here --- is the main factor limiting the trails' sizes studied with flatPERM. For instance, by not fixing any of the model parameters ($\omega_2$,  $\omega_3$ and $\kappa$), in principle we can numerically obtain the full density of states and from this the entire thermodynamic behavior of the model. It turns out, however, that with this 3-parameter approach we are not able to obtain accurate estimates of $C^{(n)}_{\{m_2,m_3,m_s\}}$ for lengths larger than $n\approx64$ steps in a reasonable amount of time. We can overcome this limitation by fixing one of the parameters and running a 2-parameter flatPERM simulation for several slices of the full parameter space. In the following, we will use this strategy to build up the finite-size phase diagram of the model for trails with up to $n=128$ steps. In order to extrapolate from these relatively short finite-size results to the phase diagram in the thermodynamic limit, we perform 1-parameter flatPERM simulations (by keeping two parameters fixed) for lengths up to $n=1024$ for values of particular interest. We also ran PERM simulations for $n\leqslant 4096$ with all parameters fixed. In all cases, results from averages of at least of $10^9$ trails are presented.

\section{Results for the Boundary Planes}
\label{boundary}

As implicit in the model definition above, we are interested here in investigating only the most physically interesting situation where all interactions are attractive. This means that we will explore the three-parameter space $(\omega_2,\omega_3,\kappa)$ where the three Boltzmann weights are larger than or equal one. To begin it is interesting to analyze the three planes forming the boundary of such space. Namely, the planes $(\omega_2,\omega_3,1)$, $(1,\omega_3,\kappa)$, and $(\omega_2,1,\kappa)$. These planes correspond to the absence of surface, two-body and three-body interactions, respectively.

\subsection{Plane $(\omega_2,\omega_3,1)$}
\label{SecPlane23}

First we investigate the case where the surface is inert, that is, it does not interact with the polymer. This should be directly comparable to the bulk case previously studied without a surface \cite{Doukas2010}. To determine the thermodynamic behavior of this plane, we analyze the density of states $\sum_{m_s}C_{\{m_2,m_3,m_s\}}$ obtained from a 2-parameter flatPERM simulation at $\kappa=1$ for $n=128$. We calculate the covariance matrix,
\begin{equation}
\begin{bmatrix} 
\left<m_2^2\right> - \left<m_2\right>^2 & \left<m_2 m_3\right> - \left<m_2\right>\left<m_3\right>\\
\left<m_3 m_2\right> - \left<m_3\right>\left<m_2\right> & \left<m_3^2\right> - \left<m_3\right>^2
\end{bmatrix}\;,
\label{Mx}
\end{equation}
whose maximum eigenvalue, $\lambda$, is a measure of the fluctuations, which can be plotted against $\omega_2$ and $\omega_3$ as a density plot. In general, local maxima in such fluctuation maps indicate the possible existence of phase boundaries there, so that we can associate lines of suitably defined maxima with pseudo-transition lines, for the finite system. Then, by considering other quantities such as the distributions of monomer contacts, we can identify the phases present in each region of the fluctuation map, and also whether the pseudo-transition lines can be associated with discontinuous or continuous phase transitions.  This method allows us to build a ``finite-size phase diagram'', which in most cases captures the main features of the actual phase diagram in the thermodynamic limit. We caution that this method does not necessarily pick up higher order phase transitions and would also identify smooth crossovers as transitions. Hence, scaling in length needs to be performed to conclusively identify the presence of phase transitions.

\begin{figure}[!t]
\includegraphics[width=7.5cm]{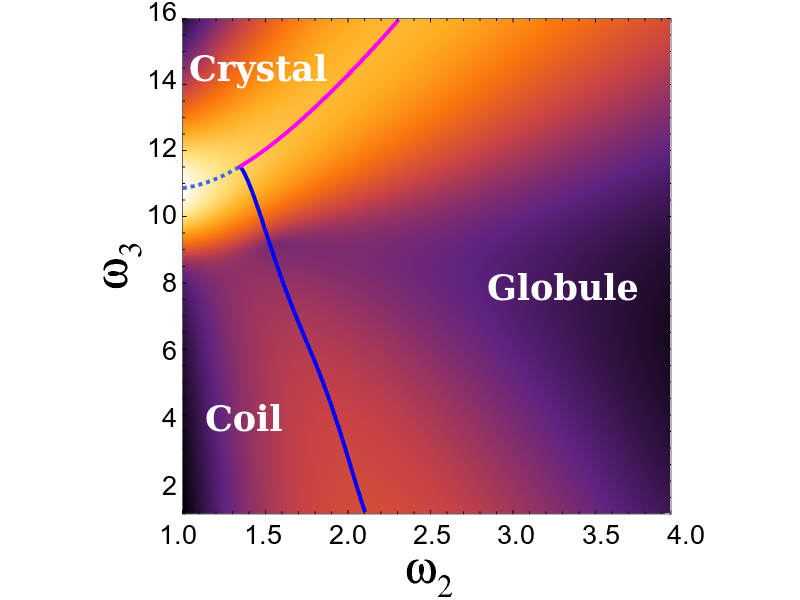}
\caption{Fluctuation map for the plane $(\omega_2,\omega_3,1)$. The lighter (darker) colors indicates regions of larger (smaller) fluctuations. The lower (higher) solid lines are approximations for the continuous coil-globule and crystal-globule transition lines, while the dashed line is the discontinuous coil-crystal transition line.}
\label{Fig2}
\end{figure}

The fluctuation map along with approximate transition lines (i.e., the finite-size phase diagram) for the plane $(\omega_2,\omega_3,1)$ is shown in Fig.~\ref{Fig2}. Distributions of doubly and triply visited sites [not shown] reveal, for small values of $\omega_2$ and $\omega_3$, the presence of configurations typically dominated by a high number of singly visited sites, characteristic of the swollen coil phase. In the region of small $\omega_3$ and large $\omega_2$, we find configurations with a large number of doubly visited sites, which may be identified as the trails' collapsed globule phase. Conversely, for small $\omega_2$ and large $\omega_3$ the trails are dominated by triply visited sites. In this region the configurations are maximally dense, so that we will refer to this phase as a crystal phase \cite{Doukas2010}. (More correctly, one should think of this phase as crystal-like, as it has a non-vanishing entropy.)

Along the line of maximal fluctuation separating the coil-globule and globule-crystal phases the distributions have a single peak, indicating the existence of continuous transitions there. Between the coil and crystal phases, on the other hand, the distributions of triply visited sites are double-peaked, suggesting that the coil-crystal transition is discontinuous. It is important to remark that it is difficult to accurately determine the transition lines when the regions of maximal fluctuation separating different phases become close to each other. For instance, in Fig.~\ref{Fig2}, we can identify only a region where the coil-crystal line change to a globule-crystal one and due to the proximity of this region with the coil-globule line, it is reasonable to infer that exists a multicritical point there, where the three lines meets. As prefigured above, such phase behavior is very similar to the one reported in \cite{Doukas2010}, for the case where the surface is absent, where indeed such lines meet at a multicritical point. This is quite expected, since the presence of an inert surface should not change the thermodynamic behavior of such systems. 

\begin{figure}[!t]
\includegraphics[angle=0, width=8.5cm]{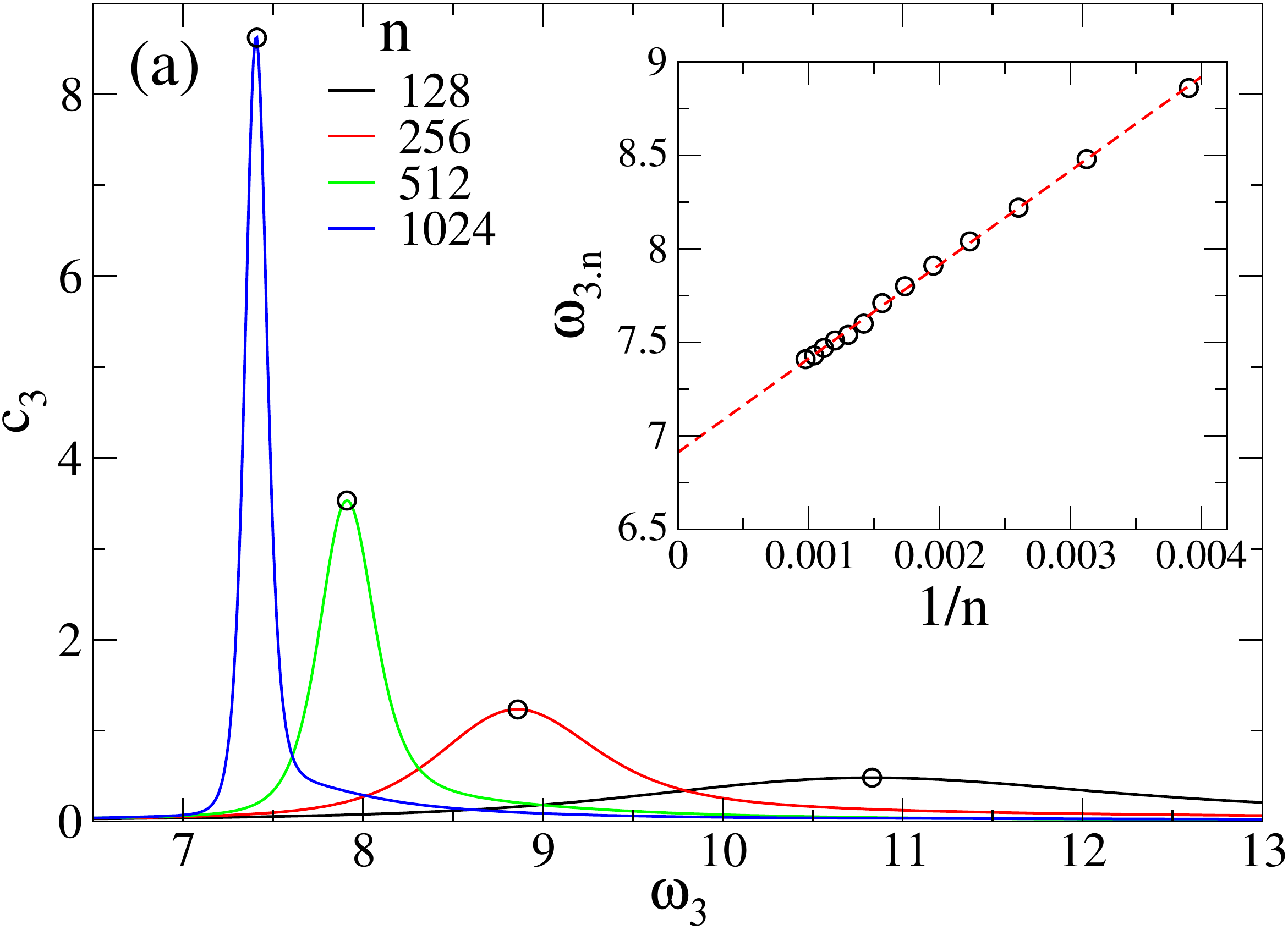}
\includegraphics[angle=0, width=8.5cm]{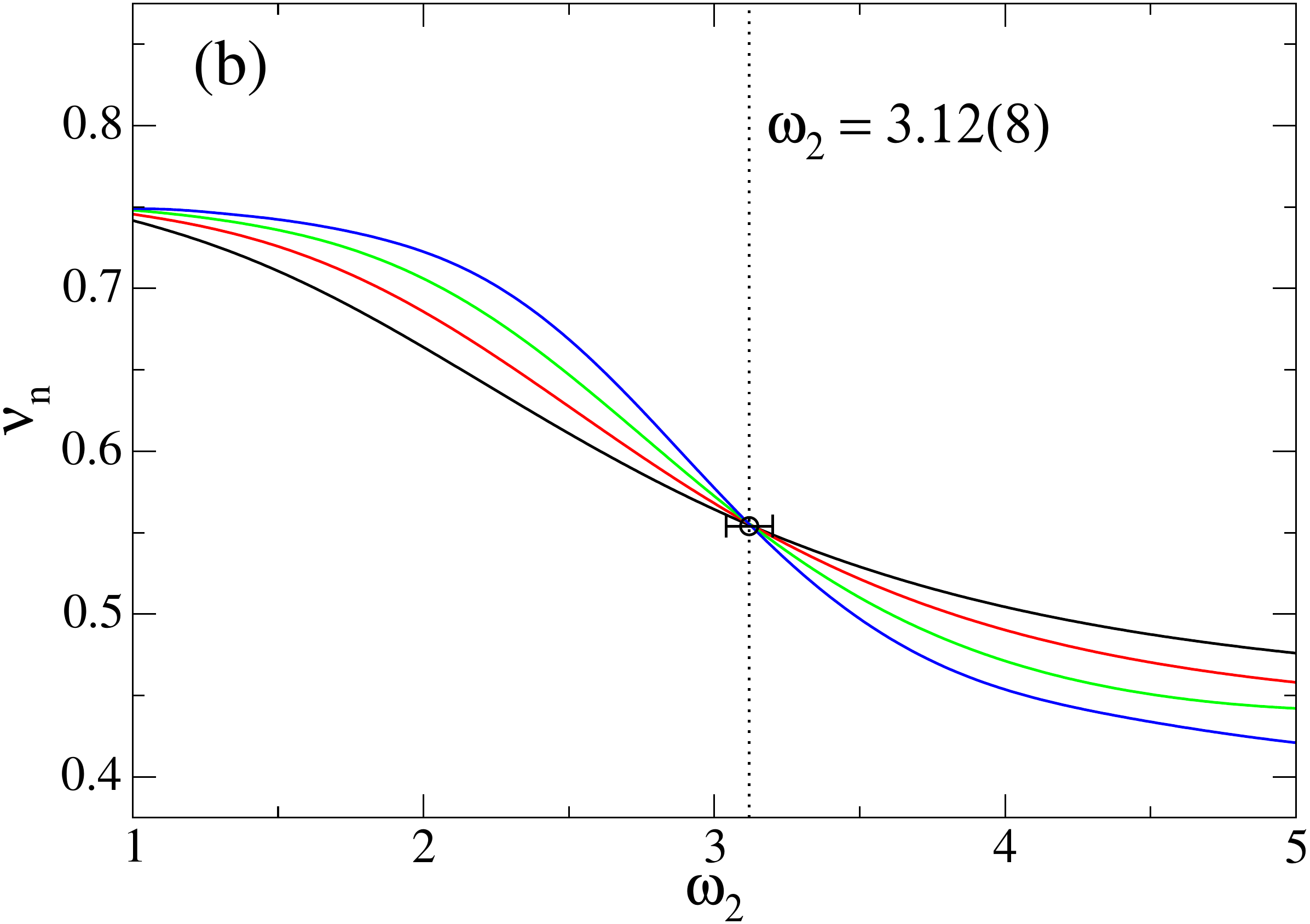}
\caption{(a) Fluctuation in the number of triply visited sites $c_3$ versus $\omega_3$, for $\omega_2=\kappa=1$ and different lengths. The open black circles indicate the maximum of each curve. In the insert the values of $\omega_{3,n}$ at the maxima are plotted against $1/n$. The dashed line is a linear fit used in the extrapolation. (b) Effective exponents $\nu_n$ as function of $\omega_2$, for $\omega_3=\kappa=1$ and several lengths. The dashed vertical line indicates the asymptotic value estimated from the crossing points.}
\label{Fig3}
\end{figure}

In order to verify the reliability of these results for small trails (with $n=128$ steps), we perform a finite-size analysis considering trails with lengths up to $n=1024$. The bulk transitions are investigated, in most cases, through the fluctuations $c_j$ in the number of doubly ($j=2$) and triply ($j=3$) visited sites. As an example, curves of $c_3$ versus $\omega_3$ are depicted in Fig.~\ref{Fig3}(a), for $\omega_2=1$ and different lengths. From the maximum in such curves, $c_{3,max}^{(n)}$, and the scaling behavior from Eqn.~\ref{BulkF}, we estimate the crossover exponent $\phi_b$, which is slightly larger than 1 in this case, consistently with the strong increase of fluctuations observed in Fig.~\ref{Fig3}(a). Then, we identify the values of $\omega_3$ at each maximum as the finite-size transition parameter and by assuming that they follow Eqn.~\ref{Tn}, with the $\phi_b$ just estimated, we extrapolate such values to obtain the transition point in the thermodynamic limit [see the insertion in Fig.~\ref{Fig3}(a)]. For $\omega_2=1$, this yields $\omega_{3}=6.91(5)$, which is on the locus of the coil-crystal transition. Similar to the findings for $n=128$ around the transition point the distribution of the number of triply visited sites still exhibits a bimodal behavior for the long trails, which together with the large $\phi_b$ exponent indicates that such transition is most likely discontinuous. 

\begin{figure}[!t]
\includegraphics[angle=0, width=8.5cm]{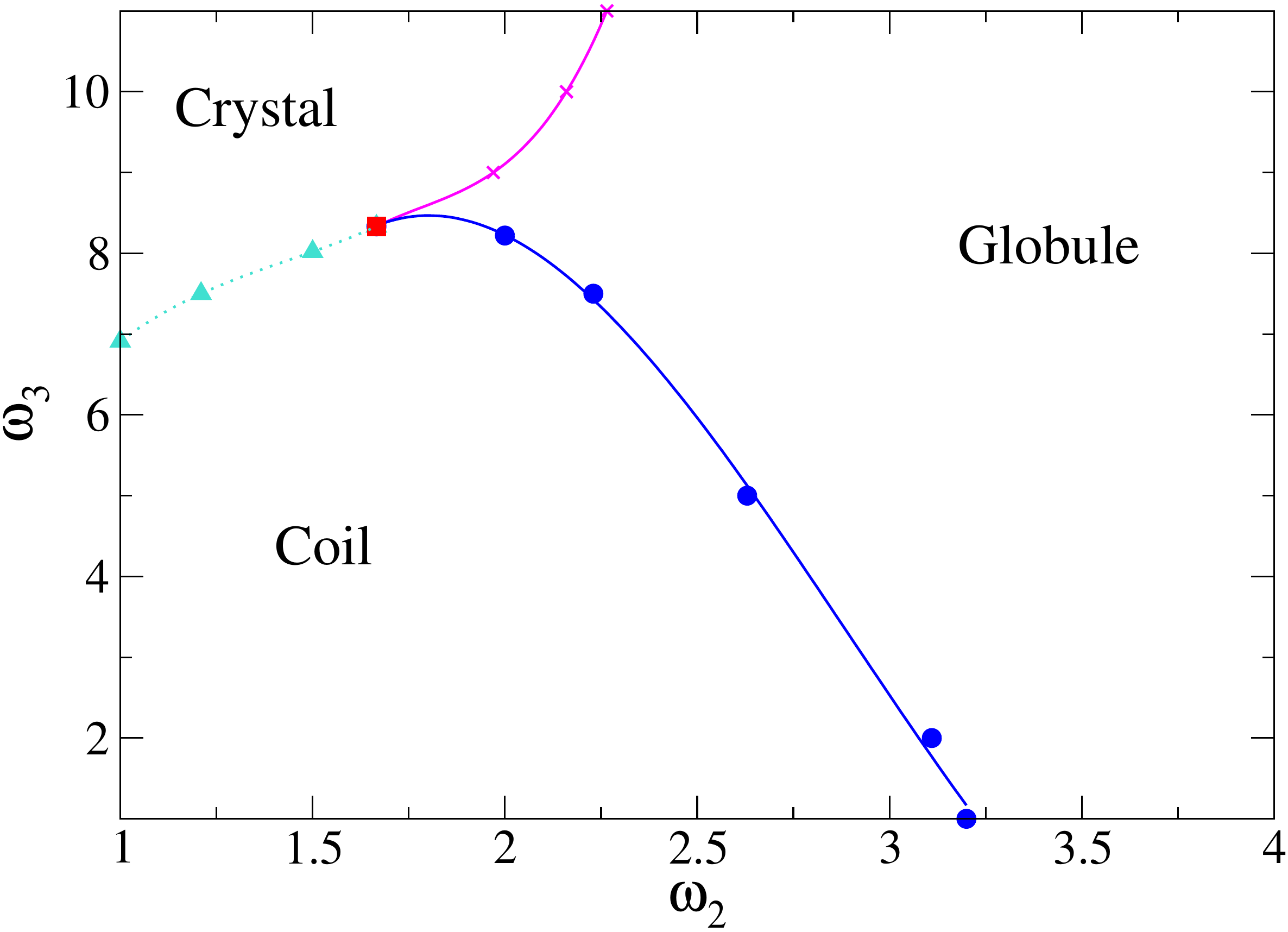}
\caption{Phase diagram for the plane $(\omega_2,\omega_3,1)$. The blue circles indicate the continuous coil-globule transition, the magenta stars the continuous globule-crystal transition and the cyan triangles the discontinuous coil-crystal transition. The red square is a multicritical point. In all cases, the lines are interpolations.}
\label{Fig4}
\end{figure}

By increasing $\omega_2$ one still finds a very similar behavior until $\omega_2 \lesssim \frac{5}{3}$, confirming the existence of a line of discontinuous coil-crystal transition in the phase diagram (see Fig.~\ref{Fig4}). Exactly at $\omega_2 = \frac{5}{3}$, we observe a continuous transition at $\omega_3=8.37(4)$. All these results are consistent with those reported by Doukas \textit{et al.} \cite{Doukas2010}, where a discontinuous coil-crystal transition ending at a multicritical point located at $(\omega_2,\omega_3) = \left(\frac{5}{3},\frac{25}{3}\right)$ was found. 

From Fig.~\ref{Fig2} and also from the results in Doukas {\it et al.}~\cite{Doukas2010}, we expect to find a coil-globule transition for (relatively) small $\omega_3$. Following a procedure similar to the one above for the coil-crystal transition, the exponent $\phi_b$ was estimated from the scaling behavior of $c_{2,max}^{(n)}$. For example, for $\omega_3=1$, one obtains $\phi_b=0.53(4)$, which is a bit larger than the value for the $\theta$ class in 2D: $\phi_{b,\theta}=3/7$ \cite{Duplantier1987}. However, as extensively reported elsewhere \cite{Lee2011,Chang1993,Meirovitch1989a}, estimates of $\phi_b$ from the bulk fluctuations are usually hampered by strong finite-size effects. For instance, by extrapolating the finite-size transition points, $\omega_{2,n}$, obtained from the maximum of $c_{2}^{(n)}$ curves for $\omega_3=1$ using $\phi_{b,\theta}$, as done in Doukas {\it et al.}~\cite{Doukas2010}, we find $\omega_2 \approx 2.6$ in agreement with the value found in \cite{Doukas2010}. However, curves the of $\omega_{2,n}$ versus $n^{-\phi_{b}}$ do not display a good linear behavior neither for $\phi_b=\phi_{b,\theta}$ nor for $\phi_b = 0.53$, indicating that the such estimate of $\omega_2$ is not so reliable. Other evidence of this is the fact that if one uses the maximum in the fluctuations in the number of trimers, instead of dimers ---  that is, $c_{3}^{(n)}$ rather than $c_{2}^{(n)}$ --- to determine $\omega_{2,n}$, an extrapolated value $\omega_2 \gtrsim 3$ is obtained (for $\omega_3=1$).

To obtain a more accurate estimate of the location of the coil-globule transition, we study the crossing of curves of effective exponents $\nu_n$ and $\gamma_n$ versus the appropriate parameter. Fig.~\ref{Fig3}(b) shows an example of this for $\omega_3=1$, where one sees that curves for a broad range of lengths cross each other in a narrow interval of $\omega_2$, demonstrating that finite-size effects are very small in this measure. In fact, by using the crossing points for pairs of successive lengths as estimates of $\omega_{2,n}$, one observes only a mild dependence with $n$. Extrapolating such values using Eqn.~\ref{Tn}, with the exponent $\phi_{b,\theta}$, we obtain $\omega_2=3.12(8)$. A very similar value is obtained from an analysis of the crossing points of effective entropic $\gamma$ exponents, where finite-size corrections are also weak. The same scenario discussed here for $\omega_3=1$ is found for larger values of $\omega_3$, along the entire coil-globule transition line. Since the results coming from the effective exponents are more accurate than those for $c_{2}^{(n)}$ and somewhat agree with those from $c_{3}^{(n)}$, we use them to determine the correct coil-globule transition line, which is shown in Fig.~\ref{Fig4}. We notice that the location of this line differs considerably from the one reported in Doukas {\it et al.}~\cite{Doukas2010}. Regardless of the numerical difference, both line estimates end at the same multicritical point, limiting similar regions of the phase diagram. Moreover, our analysis of exponents confirms the claim in \cite{Doukas2010} that this is a tricritical $\theta$-line, since the values of $\nu$ and $\gamma$ at the crossing points are found to be reasonably close to the $\theta$ exponents.

Finally, for large values of $\omega_3$ we find a continuous crystal-globule transition line [see Fig. \ref{Fig4}], which was determined following the same procedure employed above using the maxima of fluctuations. We notice that, similarly to the coil-crystal transition, here results from $c_2^{(n)}$ and $c_3^{(n)}$ are consistent. However, once again the locus of our transition line is shifted from that reported in \cite{Doukas2010}, resulting from differing finite-size analyses. Importantly, similar to the phase diagram from \cite{Doukas2010} our line seems to start at the multicritical point and extends to large values of $\omega$, as also suggested by the fluctuation map for $n=128$. Actually, despite their quantitative differences, the entire ``finite-size diagram'' from the fluctuation map in Fig.~\ref{Fig2} is consistent with the phase diagram given here in Fig.~\ref{Fig4}.

\subsection{Plane $(1,\omega_3,\kappa)$}
\label{SecPlane3k}

\begin{figure}[!b]
\includegraphics[width=8.5cm]{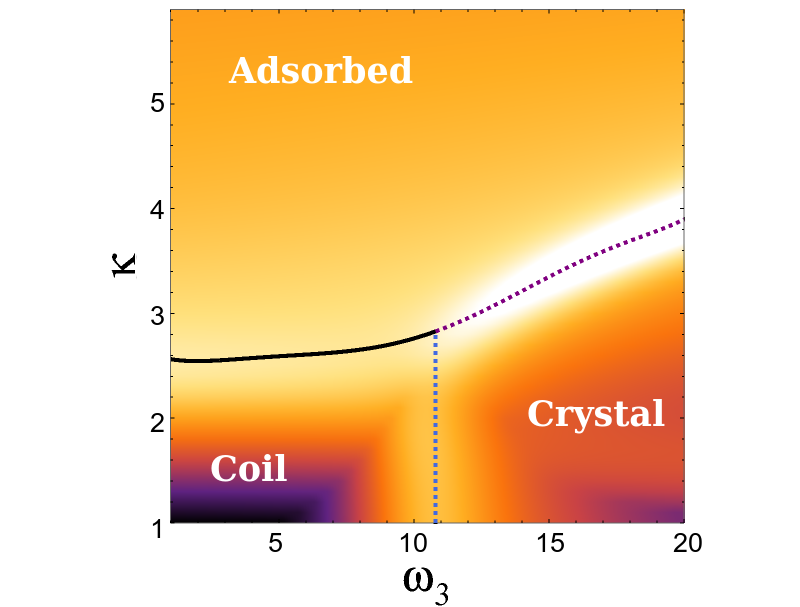}
\caption{Fluctuation map for the plane $(1,\omega_3,\kappa)$. The lighter (darker) colors indicate regions of larger (smaller) fluctuations. The solid (black) line is the continuous coil-adsorbed line, while the slanted (indigo) and vertical (blue) dashed lines are the discontinuous crystal-adsorbed and coil-crystal transition lines, respectively.}
\label{Fig5}
\end{figure}

Now, we investigate the case where the two-body monomer-monomer interaction is absent, so that $\omega_2=1$. Similar to the plane $(\omega_2,\omega_3,1)$, we will start analyzing the properties of the fluctuation map and the respective finite-size phase diagram, for $n=128$. This is shown in Fig.~\ref{Fig5}, where one clearly sees three regions separated by stripes of large fluctuations. The distributions of number of monomers reveal that such regions are associated with the phases: coil (for small $\kappa$ and $\omega_3$), crystal (for small $\kappa$ and large $\omega_3$) and adsorbed (for large $\kappa$). This last one is characterized by a high number of contacts with surface. An analysis of the modality of the distribution for the number of trimers points out that the coil-crystal and crystal-adsorbed phases are separated by discontinuous transitions, while a continuous transition line seems to exist between coil and adsorbed phases. As it is well known \cite{Vrbova1996}, when the strength of the polymer-surface interaction is not so large, the bulk transitions should not be affected by such interaction. This means that the discontinuous coil-crystal transition is expected to happen at a straight line parallel to the $\kappa$-axis in the ($\kappa,\omega_3$) space. This important observation facilitates the determination of the phase boundaries in the finite-size diagram. The maxima of fluctuations in Fig.~\ref{Fig5} suggest that the three transition lines meet at single point, which turns out to be a critical-end-point (CEP).

\begin{figure}[!t]
\includegraphics[angle=0, width=8.5cm]{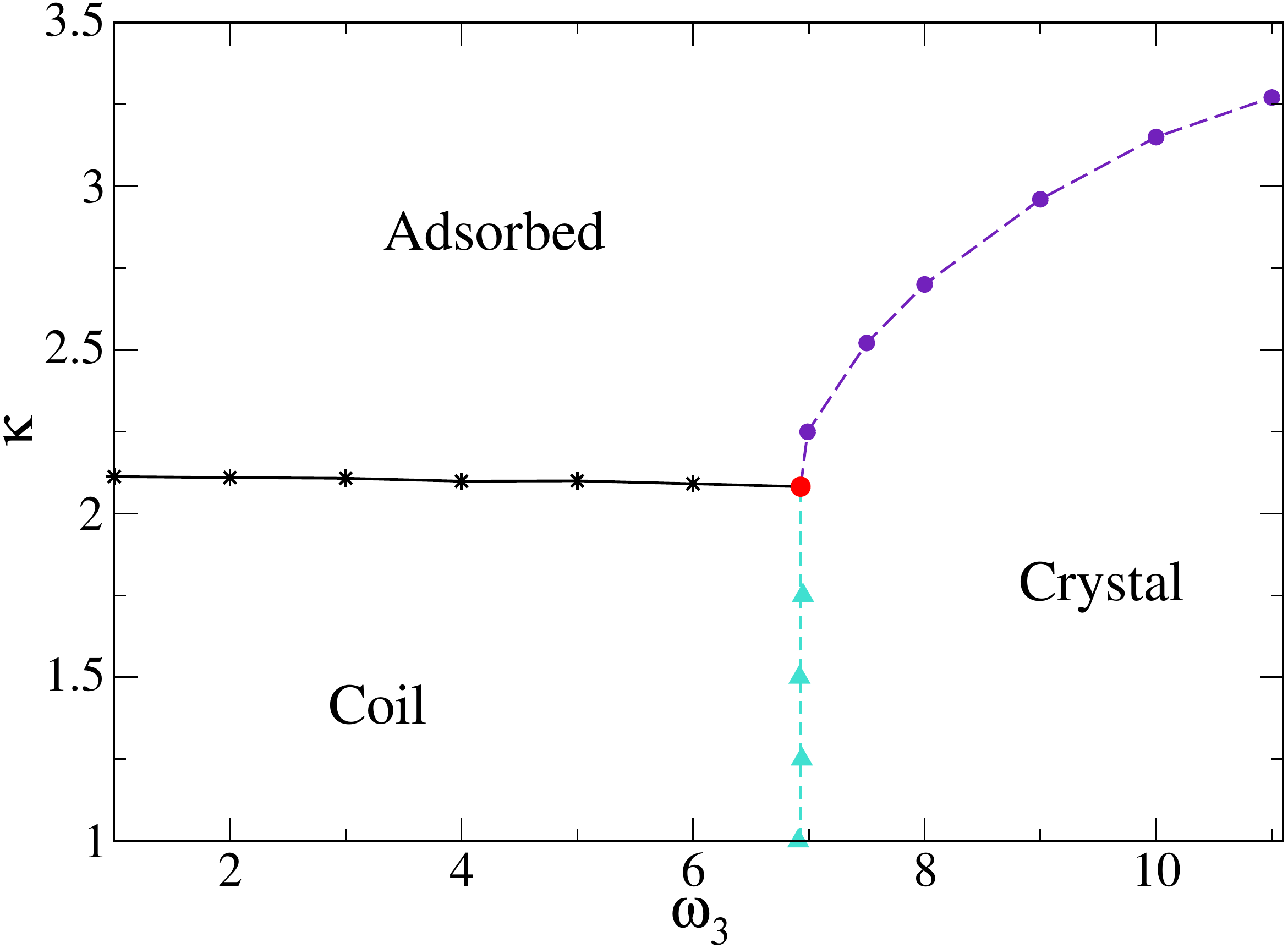}
\caption{Phase diagram for the plane $(1,\omega_3,\kappa)$. The cyan triangles indicate the discontinuous coil-crystal transition, the indigo circles the discontinuous crystal-adsorbed transition and the black stars the continuous coil-adsorbed transition. The red circle represents the critical-end-point. All lines are guide-to-eye.}
\label{Fig6}
\end{figure}

Let us now explore the phase behavior in the thermodynamic limit. To determine and confirm the existence of the (bulk) coil-crystal transition line, we follow the same procedures described in the previous subsection. This yields an approximately vertical ($\kappa$-independent) line in the phase diagram, located at $\omega_3 \approx 6.93$ (see Fig.~\ref{Fig6}). This coil-crystal coexistence line ends at $\kappa \approx 2.08$, above which it gives place to a crystal-adsorbed discontinuous transition. The crystal-adsorbed coexistence line, which was determined in the same way of the coil-crystal one, increases monotonically with $\omega_3$, as seen in Fig.~\ref{Fig6}.

\begin{figure}[!t]
\includegraphics[angle=0, width=8.5cm]{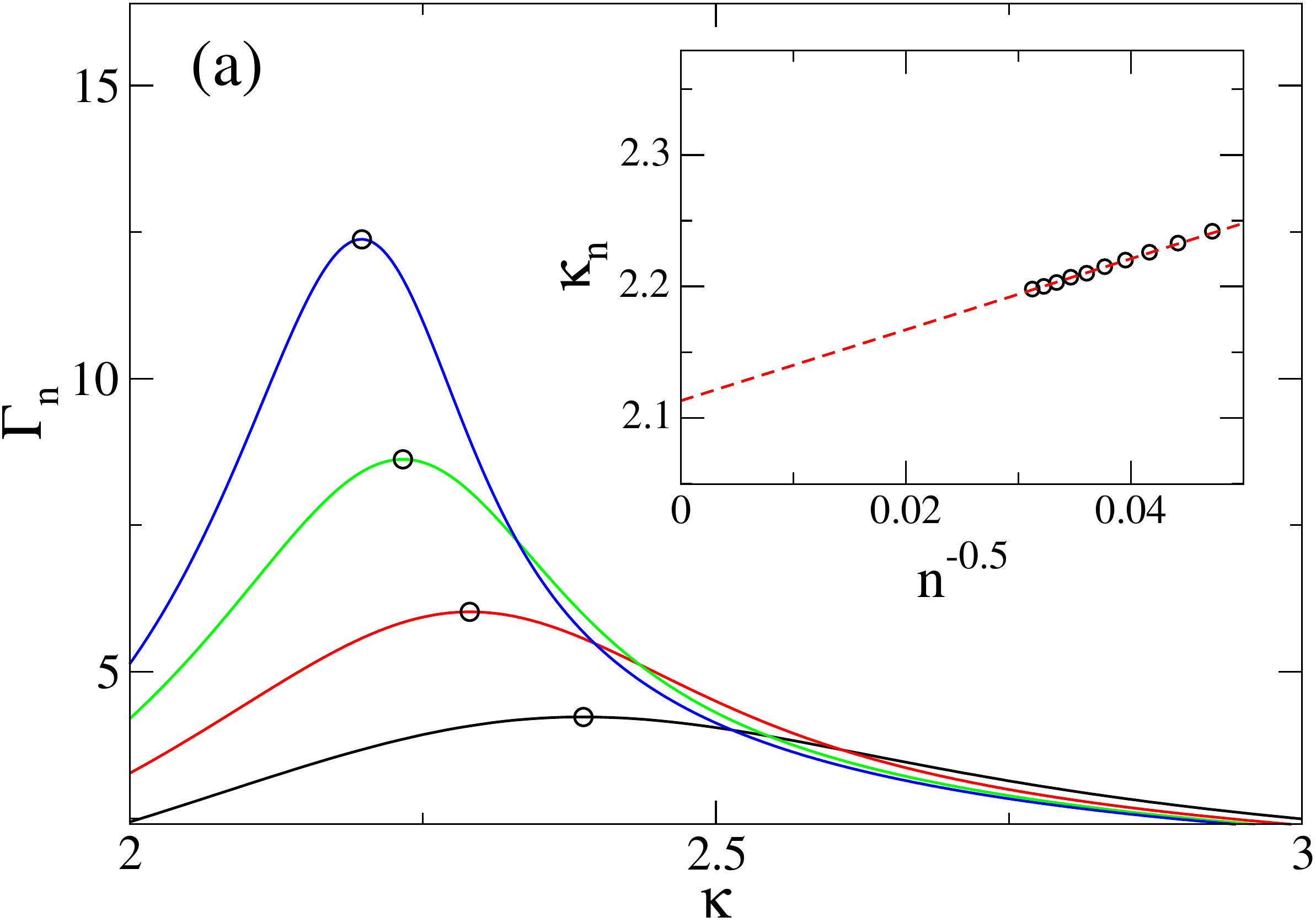}
\includegraphics[angle=0, width=8.5cm]{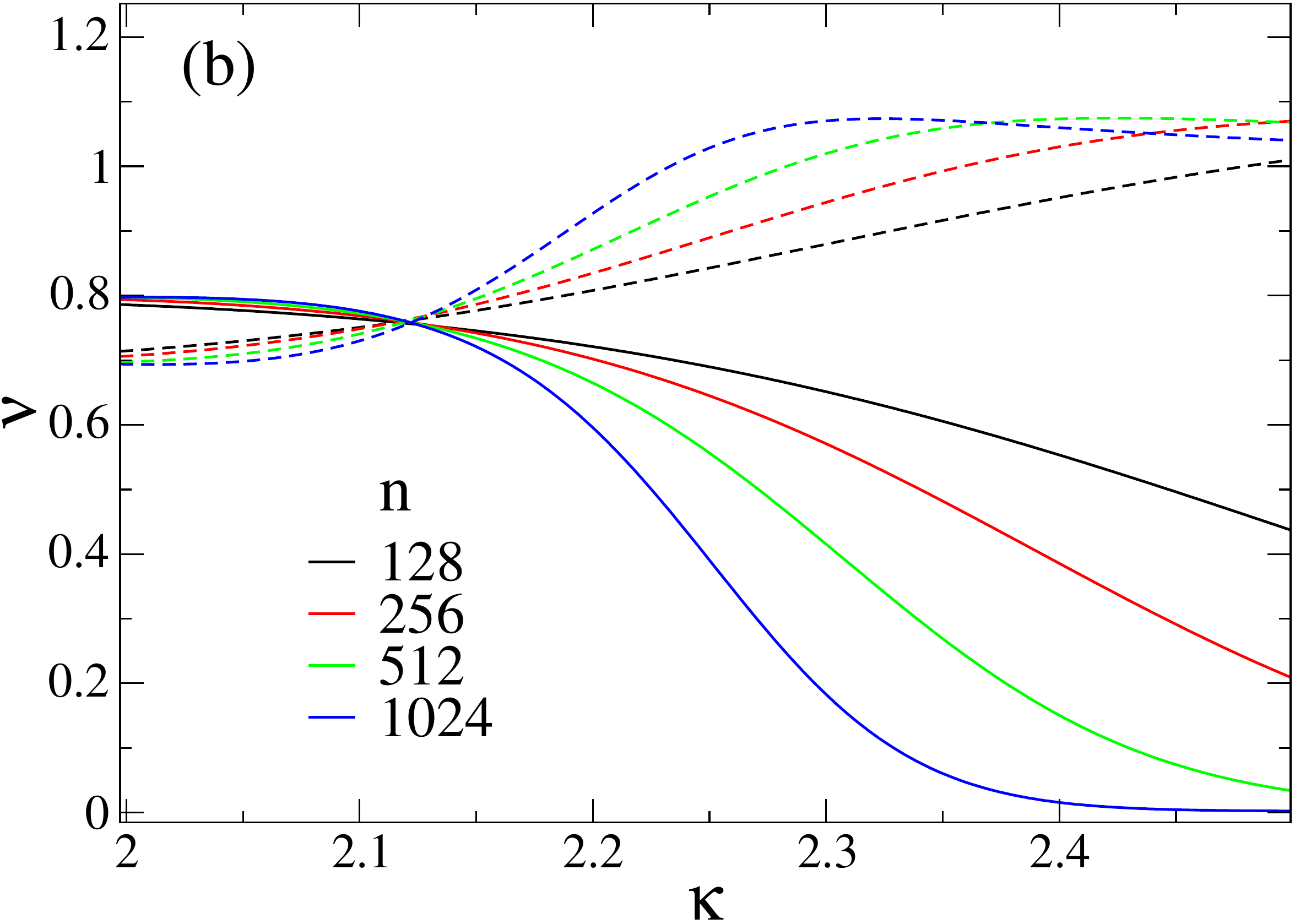}
\caption{(a) $\Gamma_n$ for different lengths as function of $\kappa$, for $\omega_2=\omega_3=1$. The circles indicate the maximum in each curve. The insertion shows the values of $\kappa_n$ at the maxima against $n^{-1/2}$. The dashed line is a linear fit used in extrapolation. (b) Effective Flory exponents $\nu_{\parallel,n}$ (dashed lines) and $\nu_{\perp,n}$ (solid lines) versus of $\kappa$, for $\omega_2=1$ and $\omega_3=2$ and several lengths.}
\label{Fig7}
\end{figure}

To analyze the (presumed continuous) coil-adsorbed transition, one starts calculating $\Gamma_n$ (Eqn.~\ref{Gn}) and using the maxima in their curves to estimate the crossover exponent $1/\delta$ from the scaling relation \ref{Gn_es}. We always find $1/\delta \approx 1/2$ along this line, in agreement with the expected value for the adsorption transition in 2D \cite{Burkhardt1989}. Then, we use the position of the maxima of $\Gamma_n$ as the finite-size transition estimate, $\kappa_n$, and extrapolate them according to Eqn.~\ref{Tn} with $1/\delta=1/2$. An example of this is shown in Fig.~\ref{Fig7}(a), for $\omega_3=\omega_2=1$, for which one obtains $\kappa=2.113(5)$.

The other accurate way of locating the adsorption transition is through the crossing of the Flory exponents $\nu_{\parallel}$ and $\nu_{\perp}$ \cite{Bradly2018}. Figure~\ref{Fig7}(b) presents an example for $\omega_3=2$. The crossing points for different lengths serve also as good estimates of $\kappa_n$ and their extrapolation (with Eqn.~\ref{Tn} and $1/\delta=1/2$) yields an estimate for the thermodynamic value for the adsorption point. Though this method returns results similar to those from fluctuations, it has weaker finite-size effects, being so more reliable to determine the adsorption transition line. Such line is also depicted in the phase diagram shown in Fig.~\ref{Fig6}, where one sees that it ends at the point where the coexistence lines meet, confirming the existence of a CEP in the diagram at $(\omega_3,\kappa)\approx(6.93,2.08)$, as suggested by the finite-size fluctuation map.

\subsection{Plane $(\omega_2,1,\kappa)$}
\label{SecPlane2k}

\begin{figure}[!t]
\includegraphics[angle=0, width=8.5cm]{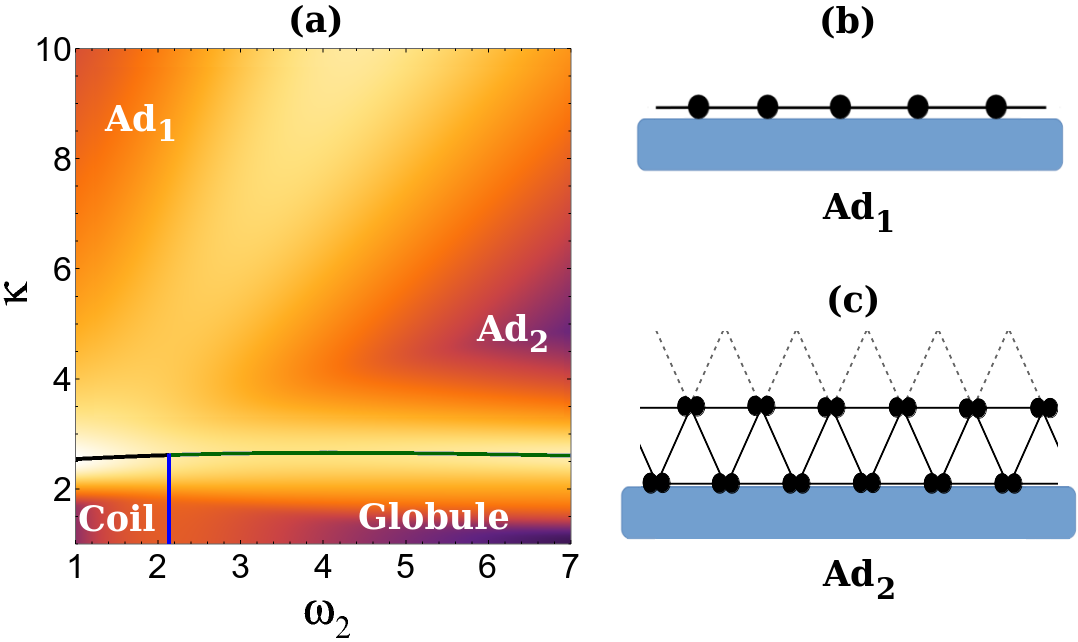}
\caption{(a) Fluctuation map for the plane $(\omega_2,1,\kappa)$. The vertical blue line indicates the coil-globule transition. The approximately horizontal line to the left (right) of the coil globule transition is the black (green) line being the coil-adsorbed (globule-adsorbed) transition line. The transitions in all these lines seem to be continuous. While the adsorbed phase is a single phase (there are no finite temperature phase transitions) it has two regions where the ground state differs. Illustrations of the two different ground state configurations are shown in (b) for the Ad$_1$ region and (c) Ad$_2$ region.}
\label{Fig8}
\end{figure}

Next, we discuss the case where trimers at a given site do not interact, so $\omega_3=1$. The fluctuation map and the finite-size transition lines are displayed in Fig.~\ref{Fig8}(a). Using the same methodology previously discussed, we find for small values of $\kappa$ indications of a continuous coil-globule transition. For large values of $\kappa$ one has signatures of two adsorbed regions. The ordinary adsorbed region --- characterized, in the ground state, by a single line of monomers lying on the surface [see Fig.~\ref{Fig8}(b)] --- is found for small $\omega_2$. For large $\omega_2$, on the other hand, we observe an adsorbed region which, in the ground state, is featured by a bilayer of doubly visited sites, as illustrated in Fig.~\ref{Fig8}(c). Hereafter, we will refer to these monolayer and bilayer regions as Ad$_1$ and Ad$_2$ regions, respectively. Importantly below our analysis indicates that there are no finite temperature transitions between them, so they are \emph{not} thermodynamically stably different phases.

Our finite-size analysis indicates that both bulk (coil and globule) phases adsorb through a continuous transition. While the globule adsorbs to the Ad$_2$ region, the coil phase adsorbs to either an Ad$_1$ or Ad$_2$ region, depending on the value of $\omega_2$. The globule adsorption to the  Ad$_2$ region can be understood, from an energetic `cost' point of view, since both feature a relatively large proportion of doubly visited sites. By locating the phase boundaries in the finite-size phase diagram, we have found indications of a possible multicritical point located where all the three continuous lines, coil-globule, coil-adsorbed and globule-adsorbed meet. 

In the fluctuation map, we see a clear region with large fluctuation between the regions Ad$_1$ and Ad$_2$, suggesting the possible existence of an Ad$_1$-Ad$_2$ transition. As already remarked, however, from the finite behavior alone we cannot determine whether this is a phase transition or simply a smooth crossover. To check this, we have performed PERM calculations for trails with up to $n=4096$ steps, for several values of $\kappa$ and $\omega_2=2.4$. We have chosen this value of $\omega_2$ since here the coil adsorbs to an Ad$_2$ region at the transition. Moreover, for very large values of $\kappa$ the adsorbed phase becomes Ad$_1$.

\begin{figure}[!t]
\includegraphics[width=8.5cm]{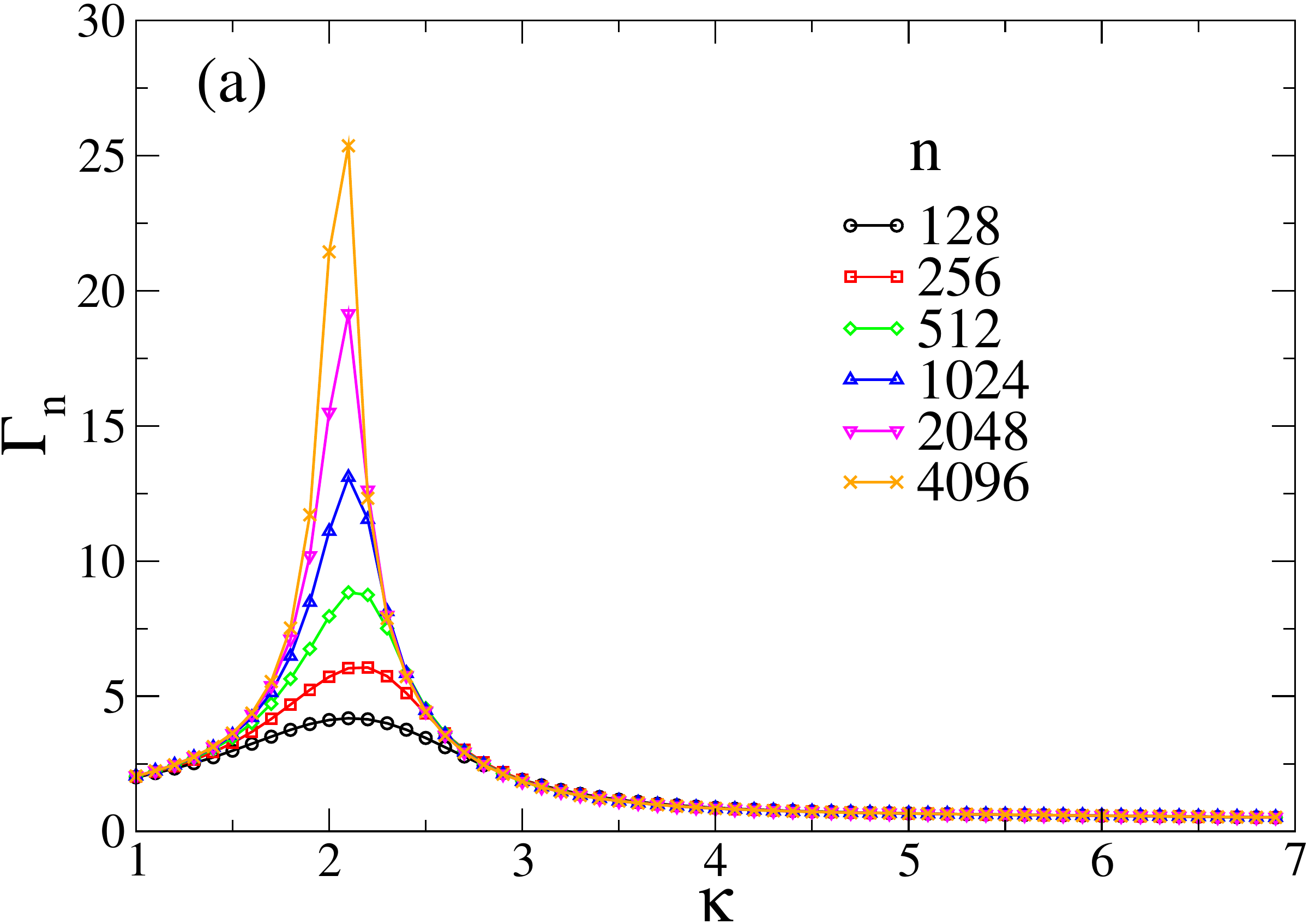}
\includegraphics[width=8.5cm]{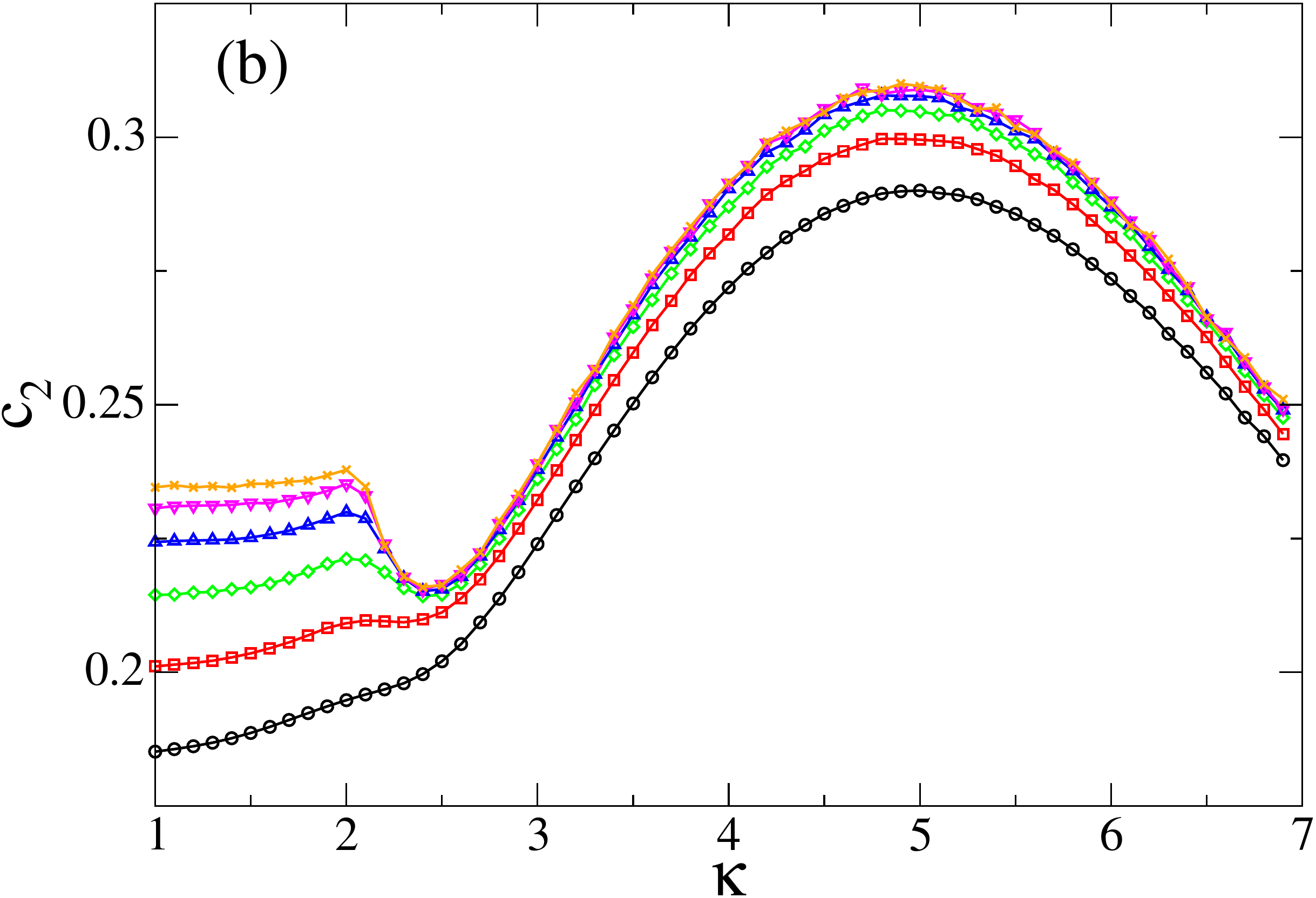}
\caption{(a) $\Gamma_n$ as function of $\kappa$ for $\omega_3=1$, $\omega_2=2.4$ and several  lengths. (b) Fluctuation in the number of doubly visited sites $c_2^{(n)}$ versus $\kappa$ for the same parameters as in (a).}
\label{Fig9}
\end{figure}

Figs.~\ref{Fig9}(a) and Fig.~\ref{Fig9}(b) present $\Gamma_n$ and $c_2^{(n)}$, respectively, from such simulations, verifying this scenario. There is a single peak in  $\Gamma_n$ around $\kappa\approx2$, where the coil-adsorbed transition is expected. There is no indication of an Ad$_1$-Ad$_2$ transition, but rather evidence for a smooth crossover.

From a physical point of view, one can understand the two regions simply as two one-dimensional layers, and short-range interactions in one-dimensional systems are not able to induce any finite-temperature phase transition, so that there can only be a smooth crossover between these regions.

\begin{figure}[!t]
\includegraphics[width=8.5cm]{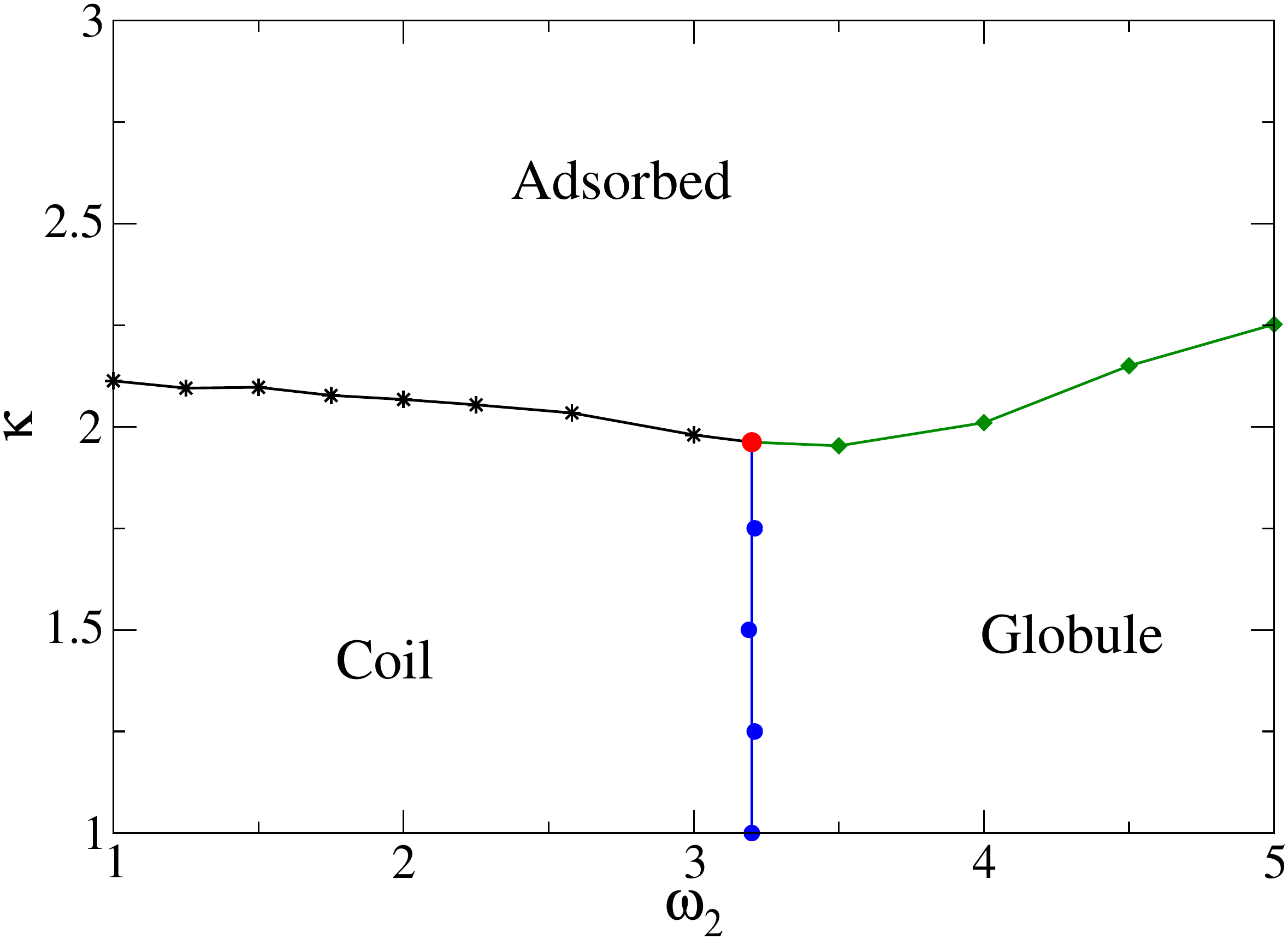}
\caption{Phase diagram for the plane $(\omega_2,1,\kappa)$. The black stars indicate the continuous coil-adsorbed transition, the blue circles the $\theta$-like coil-globule transition and the green diamonds the continuous globule-adsorbed transition. The multicritical point is denoted by the red circle. All lines are guide-to-eye.}
\label{Fig10}
\end{figure}

Therefore, by using the same methods employed in previous subsections, we obtain the phase diagram depicted in Fig.~\ref{Fig10}. Once again, it is consistent with our findings from the fluctuation map, with continuous coil-globule, coil-adsorbed and globule-adsorbed transition lines. Similarly to the coil-crystal transition found in the plane $(1,\omega_3,\kappa)$, an approximately vertical ($\kappa$-independent) straight line is found separating the coil and the globule phases, as expected. This line, located at $\omega_2\approx3.2$, starts at $\kappa=1$ and ends at a multicritical point located at $\kappa=1.96(2)$, where it meets the coil-adsorbed and globule-adsorbed lines. It is interesting to note in Fig.~\ref{Fig10} that the globule-adsorbed line appears for lower values of $\kappa$ than those at the crystal-adsorbed line (see Fig.~\ref{Fig6}). This shows that trails in the globule phase adsorb easier than those in the crystal phase. One way to understand this is that due to the presence of the Ad$_2$ region in the plane $(\omega_2,1,\kappa)$ the globule phase adsorbs to a state in which the trails can retain their large number of doubly visited sites.

\section{The three parameter model}
\label{3d}

\begin{figure}[!t]
\includegraphics[width=8.5cm]{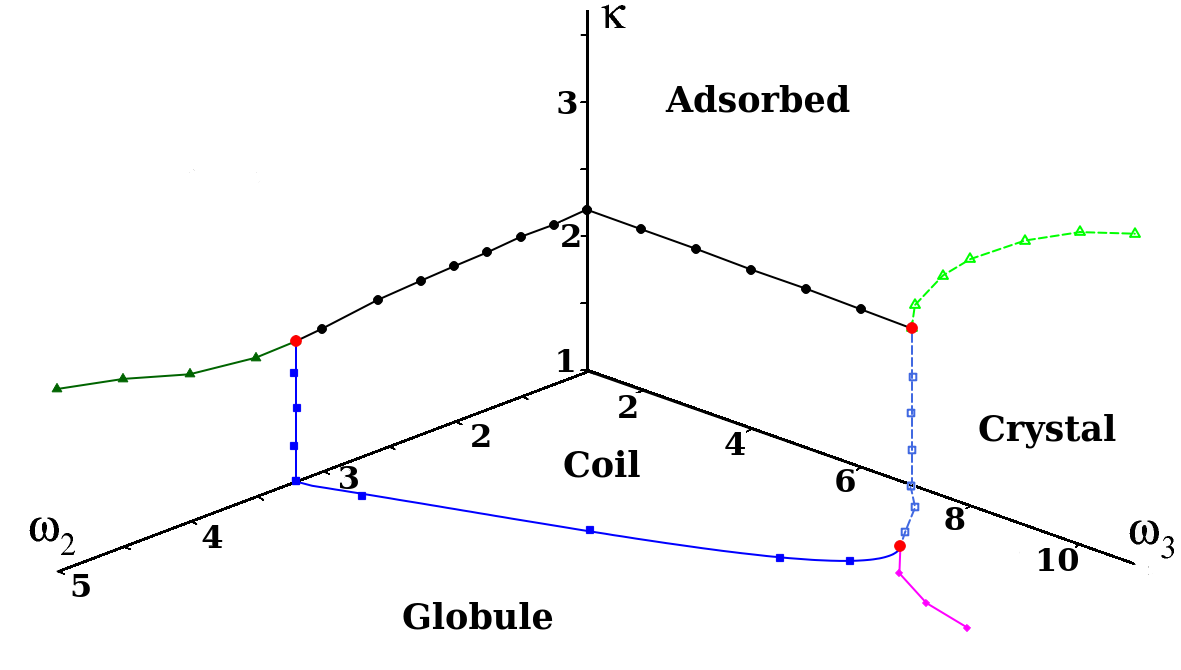}
\caption{Summary of the phases diagrams for the three boundary planes of three parameter space plotted together. The blue squares are the coil-globule transition lines estimated here. The open cyan squares are the points of discontinuous coil-crystal transitions, while the open green triangles the discontinuous crystal-adsorbed transition. The black circles denote the continuous coil-adsorbed transitions and the dark-green triangles the critical globule-adsorbed transition. The magenta diamonds indicates the continuous globule-crystal transition. The red filled circles represent the CEP and the two multicritical points.}
\label{Fig11}
\end{figure}

In order to get some insight on the phase behavior in the full three parameter space, it is instructive to begin by synthesizing the information from the three boundary planes discussed in the previous section. By simply putting these together one finds a picture as depicted in Fig.~\ref{Fig11}, which indicates the regions where each of the four thermodynamic phases (coil, globule, crystal and the adsorbed) found above appears in the full phase space and how they are separated. Of course one is assuming no new phases appear. So for instance, the continuous coil-adsorbed transition lines in the planes ($\kappa,\omega_2,1$) and ($\kappa,1,\omega_3$) strongly suggest that a critical surface exists separating these phases. Similarly, the tricritical coil-globule lines in the planes ($\kappa,\omega_2,1$) and $(\omega_2,\omega_3,1)$ indicate that such phases are separated by a tricritical (possibly $\theta$) surface. Moreover, the discontinuous transition lines between the coil and crystal phases [in the planes ($\kappa,1,\omega_3$) and $(\omega_2,\omega_3,1)$] provide strong evidence on the existence of a coil-crystal coexistence surface.

As remarked above the full three parameter flatPERM simulations could only be undertaken for length less than $n \approx 64$ to a reasonable degree of convergence. 
As such we  investigated the full phase diagram based on the finite-size behavior of two-parameter slices of the three parameter space, for $n=128$. We remark that even such 2-parameter flatPERM approach has demanded extensive simulations. As already shown in the previous section, these finite-size phase diagrams (built up from the fluctuation maps) capture the correct thermodynamic properties of the system, though quantitatively incorrect in most cases. Indeed we found no evidence of new phases, beyond the four ones already discussed. Moreover, since there is only a crossover between the adsorbed Ad$_1$ rich and Ad$_2$ rich regions, no transition line separating them will be shown, even if we indicate their locations in the figures.

\subsection{$\kappa$ slices}

\begin{figure}[!t]
\includegraphics[width=8.5cm]{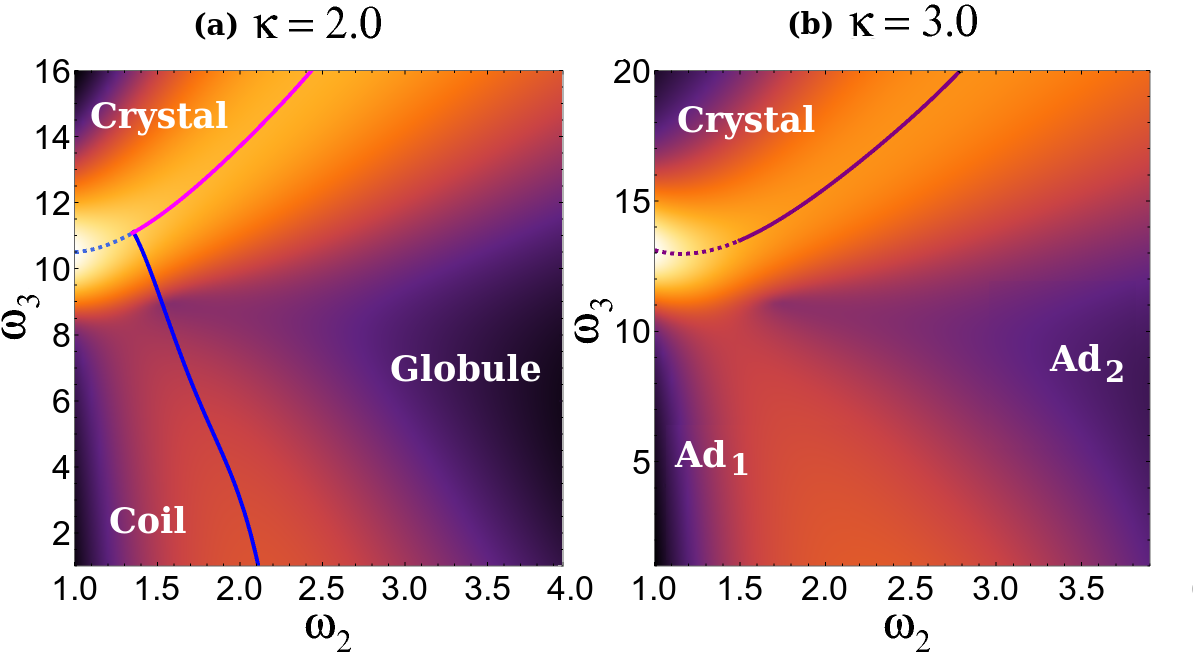}
\caption{Fluctuation maps in spaces $(\omega_2,\omega_3,2)$ and $(\omega_2,\omega_3,3)$ in (a) and (b) respectively, with transition lines indicated. In both panels solid and dashed lines indicate the presence of unimodal and bimodal energy distributions near the transition, respectively.
}
\label{Fig12}
\end{figure}

Let us start analyzing slices for $\kappa$ fixed. We recall that for $\kappa=1$ [i.e., the plane $(\omega_2,\omega_3,1)$ considered in Section \ref{SecPlane23}] the three bulk phases (coil, globule and crystal) are present in the system, as shows Figs.~\ref{Fig2}, \ref{Fig4} and \ref{Fig11}. Since, for small values of $\kappa$, the bulk transitions are not expected to be affected by the presence of an interacting surface, the same diagram of $\kappa=1$ is expected in this region. In fact, as demonstrated in Fig.~\ref{Fig12}(a), the finite-size phase diagram for $\kappa=2$ is practically identical to the one for $\kappa=1$ shown in Fig.~\ref{Fig2}. This strongly indicates that the transition lines between the three bulk phases give rise to three surfaces: a tricritical coil-globule, a critical crystal-globule and a coexistence coil-crystal, consistently with the predictions above based on Fig.~\ref{Fig11}. Moreover, the multicritical point found in the case $\kappa=1$ gives rise to a multicritical line, where the three surfaces above meet. Since this line is related to bulk phases only, we will refer to it as the bulk multicritical line (BML). Note that these bulk transition surfaces, as well as the BML are all expected to be $\kappa$-independent, i.e., perpendicular to the $(\omega_2,\omega_3)$ plane. This is indeed confirmed by the quantitative agreement between the finite-size transition lines in Figs.~\ref{Fig2} and \ref{Fig12}(a) for $\kappa=1$ and 2, respectively, as well as for intermediate $\kappa$ slices (not shown).

All these bulk transition surfaces (and the BML, as well) are expected to end when they meet the adsorption transition surfaces. Figures~\ref{Fig6}, \ref{Fig10} and \ref{Fig11} indicate that the coil-adsorbed and the globule-adsorbed critical surfaces in the thermodynamic limit change mildly with $\omega_2$ and $\omega_3$, being located close to $\kappa=2$. The same thing happens in the diagrams for $n=128$, as seen in Figs.~\ref{Fig5} and \ref{Fig8}(a), where such surfaces are in the region $2\lesssim\kappa\lesssim3$. Therefore, it is hard to observe these adsorption surfaces in planes of fixed $\kappa$. Anyhow, as shown in Fig.~\ref{Fig12}(b), for the $\kappa=3$ slice, the only bulk phase appearing in this diagram is the crystal one, confirming that we are above the coil-adsorbed and globule-adsorbed surfaces. Interestingly, in the region for small $\omega_2$ (where the coil phase was observed for smaller $\kappa$) one finds the adsorbed phase is Ad$_1$ rich, while for large values of $\omega_2$ the adsorbed phase is Ad$_2$ rich.

At finite length $n=128$, as shown in the diagram for $\kappa=3$ [Fig.~\ref{Fig12}(b)], the crystal-adsorbed transition is still found to be discontinuous for small $\omega_2$, as evidenced by a bimodal energy distribution. As discussed above, there is only one adsorbed phase, so that one would expect this transition to remain discontinuous for all values of $\omega_2$. Unfortunately, we are unable to detect a bimodal distribution for large $\omega_2$ but believe that this is simply a finite size effect.

\subsection{$\omega_2$ slices}

\begin{figure}[!b]
\includegraphics[width=8.5cm]{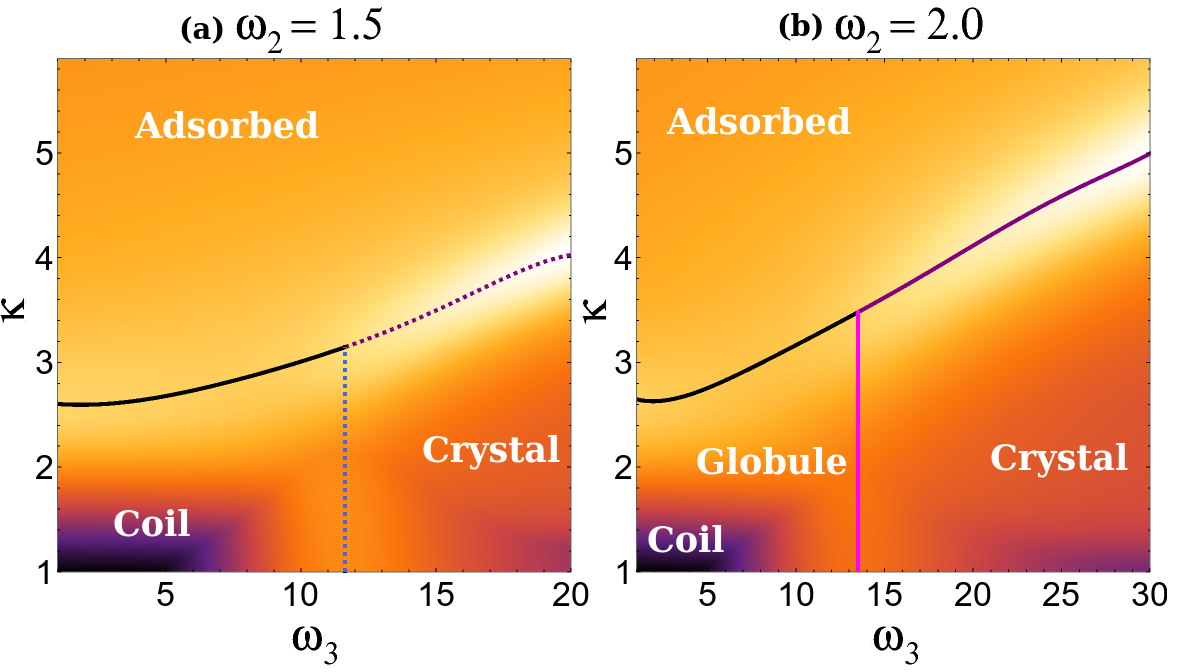}
\caption{Fluctuation maps in spaces $(1.5,\omega_3,\kappa)$ and $(2.0,\omega_3,\kappa)$ in (a) and (b) respectively, with transition lines indicated. In all panels solid and dashed lines indicate the presence of unimodal and bimodal energy distributions near the transition, respectively.
}
\label{Fig13}
\end{figure}

Now, we investigate planes for $\omega_2$ fixed. As already demonstrated in Section \ref{SecPlane3k}, for $\omega_2=1$ three phases are present in the diagram: adsorbed, coil and crystal (see Fig.~\ref{Fig5}, for $n=128$). Fig.~\ref{Fig13}(a) displays the behavior for $\omega_2=1.5$, which is the same for $\omega_2=1$, confirming the existence of a continuous coil-adsorbed transition surface, as well as of discontinuous coil-crystal and crystal-adsorbed ones. Moreover, since the former critical surface seems to end at its junction with the latter two coexistence ones, they shall form a line of critical-end-points (a CEP line) there.

The slice $\omega_2=2$ is shown in Fig.~\ref{Fig13}(b), where a different thermodynamic behavior is observed. According to Figs.~\ref{Fig2} and \ref{Fig12}, in such plane, with $\kappa$ small, one should observe transitions from coil to globule and then from globule to crystal by increasing $\omega_3$. It turns out however that the fluctuation map displayed in Fig.~\ref{Fig13}(b) fails in capturing the coil-globule transition, although for very small $\omega_3$ the trails have the characteristics of the coil phase. This caveat is certainly due to the short lengths considered. For large values of $\kappa$, the expected adsorbed phase is observed. 

We note that for the crystal-adsorbed transition at $\omega_2=2.0$ we were unable to detect a bimodal distribution. While consistent with the numerical evidence in the $\kappa$-slices we reiterate that we expect this to be a finite-size effect.

For fixed $\omega_2$, the globule-crystal transition line meets the adsorbed phase at a dense-adsorbed multicritical point, which in the full phase diagram extends to a dense-adsorbed multicritical (DAM) line. Such a line seems also to join the CEP and the BML lines at a special multicritical point. 

\subsection{$\omega_3$ slices}

\begin{figure}[!t]
\includegraphics[angle=0, width=8.5cm]{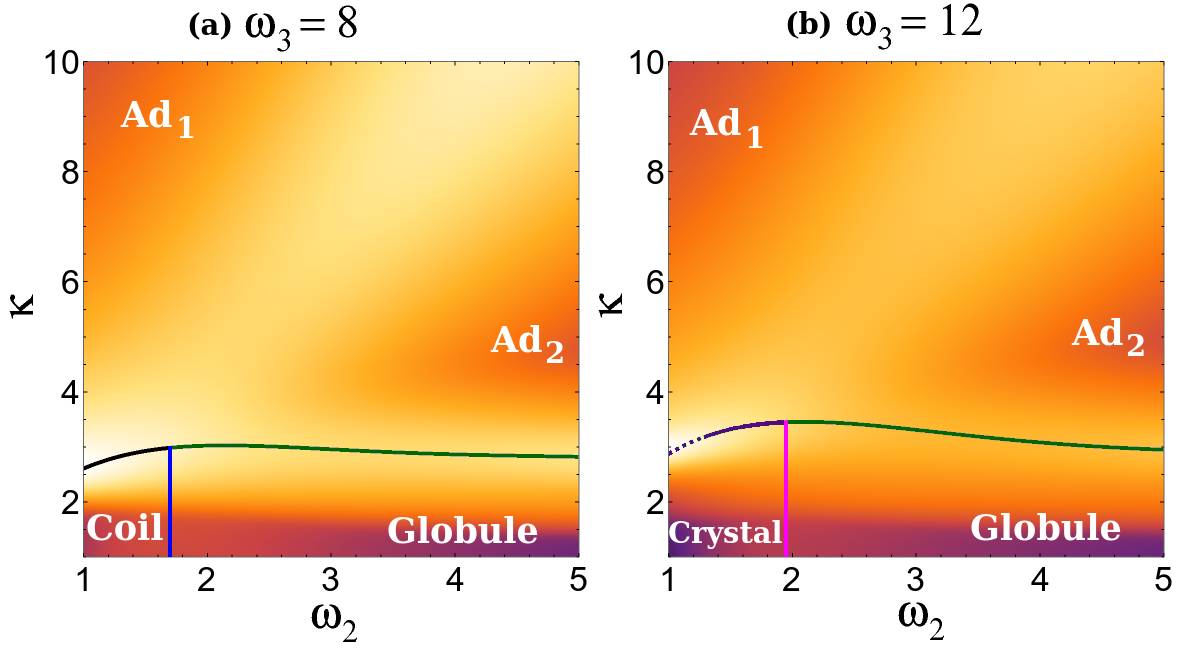}
\caption{Fluctuation maps in spaces $(\omega_2,8,\kappa)$ and $(\omega_2,12,\kappa)$ in (a) and (b) respectively, 
with transition lines indicated. In both panels solid and dashed lines indicate the presence of unimodal and bimodal energy distributions near the transition, respectively.
}
\label{Fig14}
\end{figure}

We now present the results for planes with $\omega_3$ fixed. Fig.~\ref{Fig14}(a) presents the diagram for the $\omega_3=8$ slice, which is qualitatively identical to the one for $\omega_3=1$, see Fig.~\ref{Fig8}(a); slices for intermediate values of $\omega_3$ (not shown) display the same behavior. This gives additional confirmation of the existence of a tricritical coil-globule and of critical coil-adsorbed and globule-adsorbed surfaces. These three surfaces meet, yielding a third multicritical line in the three parameter space,
which we will refer to as the collapsed-adsorbed multicritical (CAM) line. We expect that this CAM line also ends at the special multicritical point where the DAM, CEP, and BML lines meet.

As expected from Figs.~\ref{Fig2}, \ref{Fig5} and \ref{Fig11}, for large values of $\omega_3$ the crystal phase appears in place of the coil phase for small values of $\omega_2$ and $\kappa$. For $\omega_3=12$, this is demonstrated in Fig.~\ref{Fig14}(b). We are unable to detect a bimodal energy distribution along the whole crystal-adsorbed transition line, but attribute this to finite-size effects as discussed above.

\section{Summary: the full phase diagram}
\label{SecConclusions}

\begin{figure}[!t]
\includegraphics[angle=0, width=8.5cm]{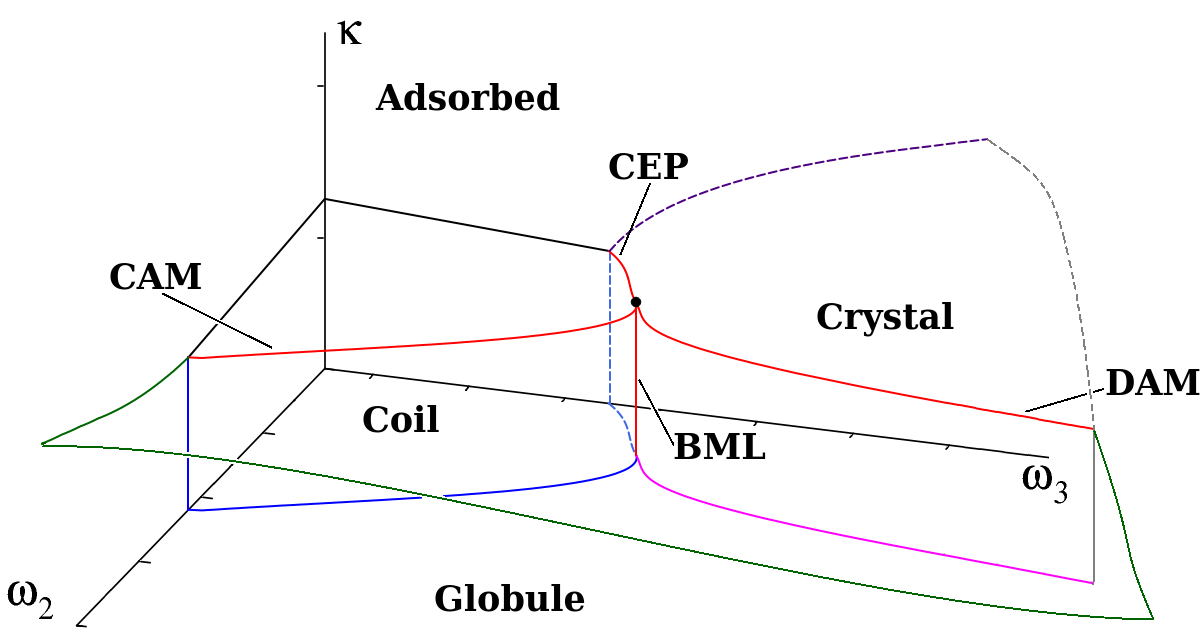}
\caption{Qualitative representation of the full phase diagram, presenting the four phases found (regarding the regions Ad$_1$ and Ad$_2$ simply as the adsorbed phase), the critical-end-point (CEP) line, as well as the bulk (BML), collapsed-adsorbed (CAM) and dense-adsorbed (DAM) multicritical lines. The special multicritical point is denoted by a circle.}
\label{Fig15}
\end{figure}

Our simulation results for self-attracting self-avoiding trails living on the triangular lattice and attractively interacting with a surface reveals a very rich thermodynamic behavior, with three bulk phases (coil, globule and crystal) and an adsorbed phase with two regions related to different ground states (the ordinary monolayer Ad$_1$ rich region and a bilayer Ad$_2$ rich region). The coil phase exists in a bounded part of the full phase diagram where the three Boltzmann weights $\omega_2,\omega_3$, and $\kappa$ are small. The crystal phase appears in the region where $\kappa$ and $\omega_2$ are relatively small, but $\omega_3$ is large. Similarly, the globule phase exists for small $\kappa$ and large $\omega_2$. Finally, the adsorbed phase appear in the region of large $\kappa$. These features are summarized in the sketch of the full phase diagram depicted in Fig.~\ref{Fig15}, where no distinction is made between both adsorbed regions, since we have determined that there is only a smooth crossover between them.

Three bulk transition surfaces are present in the system, for small $\kappa$: the continuous coil-globule (possibly in $\theta$-class \cite{Doukas2010}) and globule-crystal surfaces and a discontinuous coil-crystal surface. These three surfaces meet at the multicritical BML line, which seems to end close to $\kappa=2$. We remark that the coordinates of such line in the plane $(\omega_2,\omega_3)$ with $\kappa=1$, i.e., in the absence of the surface interaction, is exactly known to be $(\omega_2,\omega_3)=\left(\frac{5}{3},\frac{25}{3}\right)$ \cite{Doukas2010}.  Hence, by assuming that the bulk transitions are not affected by the surface interaction when $\kappa$ is small, the BML line should end at a special multicritical point located at $\left(\omega_2 =\frac{5}{3},\omega_3=\frac{25}{3},\kappa \approx 2\right)$. 

Beyond the three bulk surfaces, there are also several adsorption transition surfaces in the full phase diagram, which separate the bulk phases from the adsorbed ones. The coil-adsorbed and the globule-adsorbed are continuous, and they meet the tricritical coil-globule surface at the multicritical CAM line. 
Part of the critical coil-adsorbed surface ends at the coexistence surfaces separating the crystal phase from the coil and adsorbed ones, so that a CEP line exists there. Moreover, the globule-crystal, globule-adsorbed and crystal-adsorbed surfaces meet at another multicritical DAM line. Therefore, there are three multicritical (BML, CAM and DAM), and a CEP line 
in the phase diagram. Our results for finite trails indicate that all these lines meet at the special multicritical point discussed above.

\acknowledgments

TJO and NTR acknowledge support of the Brazilian agencies CNPq, CAPES and FAPEMIG, and the use of the computing cluster of Universidade Federal de Vi\c cosa. NTR thanks Queen Mary University London for hosting while this work was begun, and gratefully acknowledges use of the university's HPC cluster. Financial support from the Australian Research Council via its Discovery Projects scheme (DP160103562) is acknowledged by ALO.

\end{document}